\documentclass{amsart}
\usepackage{amsmath}%
\usepackage{amsfonts}%
\usepackage{amssymb}%
\usepackage{graphicx}

\newtheorem{theo}{Theorem}[section]
\newtheorem{lemme}[theo]{Lemma}

\newtheorem{coro}[theo]{Corollary}
\newtheorem{defi}[theo]{Definition}

\newtheorem{prop}[theo]{Proposition}
\newtheorem{rmq}[theo]{Remark}

\numberwithin{equation}{section}

\DeclareMathAlphabet{\mathonebb}{U}{bbold}{m}{n}
\newcommand{\one}{\ensuremath{\mathonebb{1}}}
\begin{document}
\title[Kac's equation without cutoff]{Asymptotic of grazing collisions and particle approximation for the Kac equation without cutoff.}
\author{Nicolas Fournier, David Godinho}

\begin{abstract}
The subject of this article is the Kac equation without cutoff. We first show that in the asymptotic of grazing collisions, the Kac equation can be approximated by a Fokker-Planck equation. The convergence is uniform in time and we give an explicit rate of convergence. Next, we replace the small collisions by a small diffusion term in order to approximate the solution of the Kac equation and study the resulting error. We finally build a system of stochastic particles undergoing collisions and diffusion, that we can easily simulate,  which approximates the solution of the Kac equation without cutoff. We give some estimates on the rate of convergence.
\end{abstract}
\maketitle


\vskip0.5cm

Nicolas Fournier: Laboratoire d'Analyse et de Math\'ematiques Appliqu\'ees,
CNRS UMR 8050, Universit\'e Paris-Est, 61 avenue du G\'en\'eral de Gaulle, 94010
Cr\'eteil Cedex, France

\textit{E-mail address:} nicolas.fournier@u-pec.fr
\vskip0.5cm

David Godinho: Laboratoire d'Analyse et de Math\'ematiques Appliqu\'ees,
CNRS UMR 8050, Universit\'e Paris-Est, 61 avenue du G\'en\'eral de Gaulle, 94010
Cr\'eteil Cedex, France

\textit{E-mail address:} david.godinhopereira@gmail.com
\vskip0.5cm

\textbf{Acknowledgments:} The first author was supported during this work by the grant from
the Agence Nationale de la Recherche with reference ANR-08-BLAN-0220-01.

\section{Introduction}
\subsection{The model}
The spatially homogeneous Boltzmann equation (see Cercignani \cite{CER}, Villani \cite{VIL3}) describes the density $f_t(v)$ of particles in a gas, which move with velocity $v\in\mathbb R^3$ at time $t\geq0$. The Kac equation is a one-dimensional \textit{caricature} of the Boltzmann equation. It writes

\begin{align}
\frac{\partial f_t}{\partial t}(v)=\int_{v_{*}\in \mathbb{R}}\int_{\theta=-\pi}^{\pi}\Big(f_t(v')f_t(v_{*}')-f_t(v)f_t(v_{*})\Big)\beta(\theta)d\theta dv_{*}, \label{Kac}
\end{align}
where $t\geq0$, $v\in \mathbb{R}$ and where the post-collisional velocities are given by
\begin{eqnarray}
v'=v\cos\theta-v_{*}\sin\theta,\ v_{*}'=v\sin\theta+v_{*}\cos\theta.
\end{eqnarray}
The function $\beta : [-\pi,\pi]-\{0\}\rightarrow \mathbb{R}_{+}$ is an even function called \textit{cross section}. Each pair of particles with velocities $v$ and $v_*$ collides to give particles with velocities $v'$ and $v_*'$ with a rate proportional to $\beta(\theta)$. See Kac \cite{KAC} and Desvillettes \cite{DES} for more precisions. If we have $\int_0^{\pi}\beta(\theta)d\theta= \infty$, then there is an infinite number of collisions for each particle during any time interval. The case where we assume $\int_0^{\pi}\beta(\theta)d\theta< \infty$ (case with cutoff) has been much studied. We will focus here on the real physical situation where we only assume $\int_0^{\pi}\theta^2\beta(\theta)d\theta< \infty$ (case without cutoff). By analogy with the 3d-Boltzmann equation, we will include the case where, for some $0<\nu<2$,
\begin{align} \label{beta}
\beta(\theta)\overset{\theta=0}{\approx}|\theta|^{-1-\nu}.   
\end{align}

\vskip0.5cm
We will use in this article Wasserstein distances. Let us recall that for $p\geq1$, if $f$ and $g$ are two probability measures on $\mathbb R$ with a moment of order $p$,  
$$W_p(f,g)=\inf\left\{\mathbb E(|U-V|^p)^{1/p}, U\sim f, V\sim g \right\},$$
where the infimum is taken over all random variables $U$ with law $f$ and $V$ with law $g$. See e.g. Villani \cite{VIL2} for many details on the subject. In particular, it is known that the infimum is reached : one can build $U\sim f$ and $V\sim g$ such that $W_p^p(f,g)=\mathbb E(|U-V|^p).$

\subsection{Asymptotic of grazing collisions}

Assume that there are more and more collisions, but that these collisions generate smaller and smaller deviations. For example, consider $\beta_\epsilon(\theta)=\frac{1}{\epsilon^3}\beta\Big(\frac{\pi\theta}{\epsilon}\Big)\one_{|\theta|<\epsilon}$. Then, we have $\int_0^\pi\theta^2\beta_\epsilon(\theta)d\theta=const$ and $\int_0^\pi \theta^4\beta_\epsilon(\theta)d\theta\rightarrow0.$ It is known that in this case, the solutions of Boltzmann's equation converge to the solution of the Fokker-Planck-Landau equation. To be more precise, Degond and Lucquin-Desreux \cite{DEG} and Desvillettes \cite{DES2} have shown the convergence of the operators (not of the solutions) and Villani \cite{VIL} has shown some compactness results and the convergence of subsequences. The uniqueness results of \cite{FG} show the true convergence (under some more restrictive assumptions).

If we denote by $(g_t^\epsilon)_{t\geq0}$ the solution of equation (\ref{Kac}) with cross section $\beta_\epsilon$ and initial condition $g^\epsilon_0(v)=g_0(v)$ and if we assume  $\int_{\mathbb R} v^4g_0(v)dv<\infty$  we will show that  $\sup_{t\in[0,\infty)} W_2(g_t^\epsilon,g_t)\leq	C\epsilon$, with $(g_t)_{t\geq0}$ starting from $g_0$ and solving 

\begin{equation} \label{EDPmixte}
\frac{\partial}{\partial t}g_t(v)=\frac{\mathcal{E}}{2}\frac{\partial^2}{\partial v^2}g_t(v)+\frac{1}{2}\frac{\partial}{\partial v}(vg_t(v)),
\end{equation}
where $\mathcal E:=\int_{\mathbb R} v^2g_0(v)dv$. This limit equation is \textit{nonlinear}, but the nonlinearity appears only through $\int_{\mathbb R} v^2g_t(v)dv$, which is constant in time. 

A similar result has already been proved by Toscani \cite{TOS} with a stronger distance but the rate of convergence is not very explicit. We believe that the present rate of convergence is optimal. 

\subsection{Replacing grazing collisions by a small diffusion term}

We come back to the Kac equation (\ref{Kac}) with fixed cross section $\beta$. Numerically, we must truncate small collisions, since they are in infinite number. There are two possibilities.

One may truncate roughly small collisions by replacing $\beta$ by $\tilde \beta_\epsilon(\theta)=\beta(\theta)\one_{|\theta|>\epsilon}$. We denote by $(\tilde f_t^\epsilon)_{t\geq0}$ the solution of (\ref{Kac}) with this $\tilde \beta_\epsilon.$

One may replace small collisions by a small diffusion term in the spirit of grazing collisions. We denote by $(f_t^\epsilon)_{t\geq0}$ the solution to
\begin{align} \label{mixte}
\frac{\partial}{\partial t}f_t^\epsilon(v)=&b_\epsilon\frac{\partial}{\partial v}\Big[vf_t^\epsilon(v)\Big]+\mathcal E b_\epsilon\frac{\partial^2}{\partial v^2}f_t^\epsilon(v)\\
&+\int_{v_{*}\in \mathbb{R}}\int_{|\theta|\geq\epsilon}\Big(f_t^\epsilon(v')f_t^\epsilon(v_{*}')-f_t^\epsilon(v)f_t^\epsilon(v_{*})\Big)\beta(\theta)d\theta dv_{*}, \nonumber
\end{align}
where
\begin{align} \label{beps}
b_\epsilon=\int_{|\theta|<\epsilon} (1-\cos\theta)\beta(\theta)d\theta\quad \text{and}\quad \mathcal E =\int_{\mathbb R} v^2f_0^\epsilon(v)dv.
\end{align}

We will show that $\sup_{t\in[0,T]} W_2(f_t,f_t^\epsilon)\leq C\epsilon(1+\sqrt{T})$ if $\int_{\mathbb R} v^4f_0(dv)<\infty$. Observe that when neglecting roughly grazing collisions,  we get $\sup_{t\in[0,T]} W_2(\tilde f_t^\epsilon,f_t)\leq C_T\epsilon^{1-\nu/2}$ (see Desvillettes-Graham-M\'el\'eard \cite{MEL}) if $\beta$ is as in (\ref{beta}). We can yet notice that there is no dependance on $\nu$ in our result. This is due to the fact that the more $\nu$ is close to 2, the more we neglect small collisions, but the more small collisions are well-approximated by the diffusion term. The proof is inspired by \cite{FOU}.

\subsection{A finite system of stochastic particles}

Let $\beta$ be a given cross section and $f_0$ an initial datum with $\int_{\mathbb R} v^4f_0(dv)<\infty$. We consider a solution $(f_t)_{t\geq0}$ of (\ref{Kac}).

For $\epsilon>0$ fixed, we are going to build a system of $n$ stochastic particles that we can simulate with a cost of order $Tn\int_{|\theta|>\epsilon}\beta(\theta)d\theta$ on $[0,T]$, which is at worst of order $T\epsilon^{-2}n$. If we denote by $\mu_t^{n,\epsilon}$ the empirical measure associated to this system of particles and by $\mu_t^n$ the empirical measure associated with a system of $n$ i.i.d. particles with law $f_t$, we will show that
$$\sup_{[0,T]}\mathbb E\Big[W_2^2(f_t,\mu_t^{n,\epsilon})\Big]\leq C(1+T)^3\Big(\epsilon^2+\sup_{[0,T]}\mathbb E\Big[W_2^2(f_t,\mu_t^n)\Big]\Big).$$
Our system of particles is thus as efficient as the system of particles with \textit{true i.i.d. particles} with law $f_t$ which are not simulable because of the nonlinearity. If we assume that $f_0$ has infinitely many moments, we will get
\begin{align*}
\sup_{[0,T]}\mathbb E\Big[W_2^2(f_t,\mu_t^{n,\epsilon})\Big]\leq C(1+T)^3\Big(\epsilon^2+\frac{1}{n^{(1/2)^-}}\Big).
\end{align*}

This system of particles uses the ideas of the previous section : we replace small collisions by a small diffusion term, which gives an error of order $\epsilon$.

In Desvillettes-Graham-M\'el\'eard \cite{MEL}, they just cutoff small collisions and they get, roughly, something like $\sup_{[0,T]}\mathbb E\Big[W_2^2(f_t,\tilde\mu_t^{n,\epsilon})\Big]\leq C_T\Big(\epsilon^{2-\nu}+\frac{e^{C_T\Lambda_\epsilon}}{n}\Big)$, with $\Lambda_\epsilon=\int_{|\theta|>\epsilon}\beta(\theta)d\theta\approx\epsilon^{-\nu}$ if $\beta$ is as in (\ref{beta}). If we compare this result with our result, we can observe the following.
\begin{itemize}
	\item In the first term, we get an error of order $\epsilon^2$ instead of $\epsilon^{2-\nu}$. It is due to the fact that we replace small collisions by a small diffusion term.
	\item In the second term, we get a bound which does not depend on $\epsilon$. It is because we use a Wasserstein distance which is well-adapted for this study. In Desvillettes-Graham-M\'el\'eard \cite{MEL}, they give the final result with a Wasserstein distance, but to get this result they use a variation distance.
	\item The cost of simulation for the two systems of particles is similar. 
\end{itemize}

See also Peyre \cite{PEY} who gives large deviations estimates for the Boltzmann equation for Maxwell molecules and Mischler-Mouhot \cite{MIS} who give results of chaos propagation with quantitative estimates for the Boltzmann equation for hard spheres and for Maxwell molecules.

\subsection{Comments}

We managed to obtain some bounds uniform in time for the asymptotic of grazing collisions. For our two other main results, we tried to limit the time dependance. We thus avoid getting bounds with exponential terms.
	
The bound we get for $\mathbb E[W_2^2(f_t,\mu_t^n)]$ is not very satisfactory. \textit{A priori}, it is of order $n^{-(1/2)-}$ (if the initial condition has infinitely many moments, see Lemma \ref{empirique}) which gives a bound for $\mathbb E [W_2(f_t,\mu_t^n)]$ of order $n^{-(1/4)-}$. We expected to get a bound of order $n^{-1/2}$ as in the central limit theorem, but we cannot get it. See Peyre \cite{PEY} for example to get more details. It seems to be the only defect of $W_2$ for this study. 
	
Assuming that $\int_0^\pi\theta\beta(\theta)d\theta<\infty$ (e.g. if we assume (\ref{beta}) with $\nu\in(0,1)$), we get a bound for $\mathbb E [W_1(f_t,\mu_t^{n,\epsilon})]$ which is of order $\epsilon+n^{-1/2}$ but with an exponential dependance in time. If $\int_0^\pi\theta^\gamma\beta(\theta)d\theta<\infty$ (e.g. if $\nu<\gamma$) for some $\gamma\in(1,2)$, we also study $\mathbb E[W_\gamma^\gamma(f_t,\mu_t^{n,\epsilon})]$.

In a future work, we will apply the same kind of methods to the homogeneous Boltzmann equation. We hope to get some results which will probably be much less optimal. 
	
Our proofs use probabilistic methods, which was initiated in the famous paper of Tanaka \cite{TAN}, and used in Desvillettes-Graham-M\'el\'eard \cite{MEL}. We will also use a result of Rio \cite{RIO}, which gives some very precise rate of convergence for the standard central limit theorem in Wasserstein distance.

\subsection{Plan of the paper}

In the next section, we will state more precisely our three main results. In Section 3, we will give a probabilistic interpretation of the three equations. Sections 4, 5 and 6 are devoted to the proofs of our main results. Some numerical illustrations will be given in Section 7. At the end of the paper, we will give an appendix with some results about the Wasserstein distance between a compensated Poisson integral and a centered Gaussian law with same variance, the rate of convergence of an empirical measure using Wasserstein distances, the moments of the solution to (\ref{Kac}) and the well-posedness for a certain kind of P.D.E.s.

\section{Results}

\subsection{Weak solutions}

Let $\beta$ be a cross section satisfying  
\begin{align} \label{ncutoff}
\int_{-\pi}^{\pi} \theta^2\beta(\theta)d\theta<\infty.
\end{align}
For $k\geq0$, we denote by $\mathcal{P}_k(\mathbb{R})$ the set of probability measures on $\mathbb R$ admitting a moment of order $k$ and by $C_b^2(\mathbb{R})$ the space of real bounded functions which are in $C^2(\mathbb{R})$ with first and second derivatives bounded. We say that a family of probability measures $(f_t)_{t\geq0}$ is in $L_{loc}^\infty\big([0,\infty),\mathcal P_2(\mathbb R)\big)$ if $\sup_{[0,T]} \int_{\mathbb R} v^2f_t(dv)<\infty$ for all $T$. If $\varphi\in C_b^2(\mathbb{R})$ and $(v,v_*)\in\mathbb R^2$, we set
\begin{align} \label{Kbeta}
\nonumber K_{\beta}^{\varphi}(v,v_{*})=&\int_{-\pi}^{\pi}\Big[\varphi(v\cos\theta-v_{*}\sin\theta)-\varphi(v)-(v(\cos\theta-1)-v_{*}\sin\theta)\varphi'(v)\Big]\beta(\theta)d\theta\\ 
&-bv\varphi'(v),
\end{align}
with
\begin{eqnarray}
b=\int_{-\pi}^{\pi}(1-\cos\theta)\beta(\theta)d\theta. \label{b}
\end{eqnarray}
If $\int_0^\pi \theta\beta(\theta)d\theta<\infty$, then one easily checks, using that $\beta$ is even, that
\begin{align}  K_{\beta}^{\varphi}(v,v_{*})=\int_{-\pi}^{\pi}\Big[\varphi(v\cos\theta-v_{*}\sin\theta)-\varphi(v)\Big]\beta(\theta)d\theta
\end{align}

We now define precisely the notion of solutions that we will use.

\begin{defi} \label{weak solutions}
Consider a cross section $\beta$ satisfying (\ref{ncutoff}).
\begin{enumerate}
	\item We say that $(f_t)_{t\geq0}\in L_{loc}^\infty\big([0,\infty),\mathcal P_2(\mathbb R)\big)$ solves (\ref{Kac}) if for any $\varphi$ in $C_b^2(\mathbb{R})$, any $t\geq0$,
\begin{align} 
 \int_{\mathbb R}\varphi(v) f_t(dv)=\int_{\mathbb R} \varphi(v) f_0(dv)+\int_0^t\int_{\mathbb R}\int_{\mathbb R}  K_{\beta}^{\varphi}(v,v_{*})f_s(dv)f_s(dv_{*}) ds. \label{Kac mesure} 
\end{align}
	\item We say thay $(g_t)_{t\geq0}\in L_{loc}^\infty\big([0,\infty),\mathcal P_2(\mathbb R)\big)$ solves (\ref{EDPmixte}) if for any $\varphi$ in $C_b^2(\mathbb{R})$, any $t\geq0$,
\begin{align}
\label{lim}
\int_{\mathbb R}\varphi(v) g_t(dv)=&\int_{\mathbb R} \varphi(v) g_0(dv)+\frac{1}{2}\mathcal{E}\int_0^t\int_{\mathbb R} \varphi''(v)g_s(dv) ds\\ \nonumber
&-\frac{1}{2}\int_0^t\int_{\mathbb R} v\varphi'(v)g_s(dv)ds, 
\end{align}
where $\mathcal E:=\int v^2g_0(dv)$.
	\item For $\epsilon\in(0,1)$ fixed, we say that $(f_t^\epsilon)_{t\geq0}\in L_{loc}^\infty\big([0,\infty),\mathcal P_2(\mathbb R)\big)$ solves (\ref{mixte}) if for any $\varphi$ in $C_b^2(\mathbb{R})$, any $t\geq0$,
\begin{align}
\nonumber 
\int_{\mathbb R} \varphi(v) f_t^\epsilon(dv)=&\int_{\mathbb R} \varphi(v) f_0^\epsilon(dv)-b_\epsilon\int_0^t\int_{\mathbb R} v\varphi'(v) f_s^\epsilon(dv)ds\\ \label{faible}
&+ \mathcal E b_\epsilon\int_0^t\int_{\mathbb R} \varphi''(v)f_s^\epsilon(dv) ds+\int_0^t\int_{\mathbb R}\int_{\mathbb R} K_{\beta_\epsilon}^{\varphi}(v,v_{*})f_s^\epsilon(dv)f_s^\epsilon(dv_{*}) ds, 
\end{align}
where
\begin{align} \label{beps2}
\beta_\epsilon(\theta)=\beta(\theta)\one_{|\theta|>\epsilon},\quad b_\epsilon=\int_{|\theta|<\epsilon}(1-\cos\theta)\beta(\theta)d\theta\quad and\quad \mathcal E=\int_{\mathbb R} v^2 f_0^\epsilon(dv).
\end{align}
\end{enumerate}
\end{defi}
Observe that all the terms in the above equations are well-defined. For example in (\ref{Kac mesure}), the last term is well-defined because for $\varphi\in C_b^2(\mathbb R)$, $|K_{\beta}^{\varphi}(v,v_{*})|\leq C\int_0^\pi \theta^2\beta(\theta)d\theta(|v|^2+|v_*|^2)||\varphi''||_\infty+b|v|||\varphi'||_\infty$.
\vskip0.5cm

\begin{prop} \label{ex}
Let $f_0$, $g_0$ and $f_0^\epsilon$ be in $\mathcal{P}_2(\mathbb{R})$ and let $\beta$ satisfy (\ref{ncutoff}). There is existence and uniqueness of solutions $(f_t)_{t\geq0}$, $(g_t)_{t\geq0}$ and $(f_t^\epsilon)_{t\geq0}$ to equations (\ref{Kac}), (\ref{EDPmixte}) and (\ref{mixte}) starting from $f_0$, $g_0$ and $f_0^\epsilon$ respectively, in the sense of Definition \ref{weak solutions}. Furthermore, we have energy conservation: for any $t\geq0$
\begin{align}\label{energie}
\int_{\mathbb R}v^2f_t(dv)=\int_{\mathbb R}v^2f_0(dv),\ \int_{\mathbb R}v^2g_t(dv)=\int_{\mathbb R}v^2g_0(dv)
\end{align}
and
\begin{align}
\int_{\mathbb R}v^2f_t^{\epsilon}(dv)=\int_{\mathbb R}v^2f_0^{\epsilon}(dv).
\end{align}
\end{prop}
\vskip0.5cm

For the proof of the previous result, one can see Toscani-Villani \cite{TOSVIL} for (\ref{Kac}). For (\ref{EDPmixte}), use Proposition \ref{unicite} with $a=\frac{1}{2}\mathcal E$, $b=-\frac{1}{2}$ and $q=r=0$. For (\ref{mixte}), use Proposition \ref{unicite} with $a=\mathcal Eb_\epsilon$, $b=-b_\epsilon$, $q=0$ and $r(t,v,v_*,dh)$ defined by $r(t,v,v_*,A)=\int_{-\pi}^\pi \one_{|\theta|>\epsilon}\one_{A}\big(v(\cos\theta-1)-v_*\sin\theta\big)\beta(\theta)d\theta$ for all Borel subset $A$ of $\mathbb R$, which indeed satisfies $\sup_{t,v,v_*} r(t,v,v_*,\mathbb R)=\int_{-\pi}^\pi\one_{|\theta|>\epsilon}\beta(\theta)d\theta<\infty$ and $\sup_{t\geq0}\int_{\mathbb R} (h^2+2vh)r(t,v,v_*,dh)=\int_{|\theta|>\epsilon}\sin^2\theta\beta(\theta)d\theta(v^2+v_*^2)= C(v^2+v_*^2)$. To get energy conservation, it suffices to apply (\ref{Kac mesure}), (\ref{lim}) and (\ref{faible}) with $\varphi(v)=v^2$.

\subsection{Asymptotic of grazing collisions}

Our first main result is the following.
\begin{theo} \label{collisions rasantes}
Let $g_0\in\mathcal P_4(\mathbb R)$ and let $(\beta_\epsilon)_{\epsilon\in(0,1)}$ be a family of cross sections verifying $\int_{-\pi}^{\pi}\theta^2 \beta_{\epsilon}(\theta)d\theta=1$ and $\int_{-\pi}^{\pi}\theta^4 \beta_{\epsilon}(\theta)d\theta\stackrel{\epsilon\rightarrow0}{\longrightarrow}0$. For $\epsilon\in(0,1)$, let $(g_t^\epsilon)_{t\geq0}$ be the solution of (\ref{Kac}) with $g_0$ for initial datum and $\beta_{\epsilon}$ for cross section. If $(g_t)_{t\geq0}$ is the solution of (\ref{EDPmixte}) with the same $g_0$ for initial datum, then for all $\epsilon\in(0,1)$,
\begin{align*}
\sup_{t\in[0,\infty)} W_2^2(g_t^\epsilon,g_t)\leq	C\frac{\int_{\mathbb R}v^4g_0(dv)}{\mathcal E}\int_{-\pi}^{\pi}\theta^4\beta_{\epsilon}(\theta)d\theta,
\end{align*}
where $C$ is a universal constant.
\end{theo}
\vskip0.5cm

This convergence result was already known (see for example Toscani \cite{TOS}), but we get here an explicit and probably optimal rate of convergence, which, to our knowledge, had never been done so far.

\begin{rmq}
If we consider a cross section $\beta$ with $\int_{-\pi}^{\pi}\theta^2 \beta(\theta)d\theta=1$ and if for any $\epsilon \in(0,1)$, we set  $\beta_{\epsilon}(\theta)=\frac{\pi^3}{\epsilon^3}\beta\Big(\frac{\pi\theta}{\epsilon}\Big)\one_{|\theta|<\epsilon}$, then $\int_{-\pi}^{\pi}\theta^4\beta_\epsilon(\theta)d\theta\leq	\epsilon^2.$
\end{rmq}

\subsection{Error when we replace the small collisions by a small diffusion term}
Let us explain briefly why (\ref{faible}) approximates (\ref{Kac mesure}): consider a cross section $\beta$ satisfying (\ref{ncutoff}) and $\varphi\in C_b^2(\mathbb R)$. Write, using that $\beta$ is even,
\begin{align*}
K_{\beta}^{\varphi}(v,v_{*})&=\int_{-\epsilon}^\epsilon\Big[\varphi(v\cos\theta-v_{*}\sin\theta)-\varphi(v)-(v(\cos\theta-1)-v_{*}\sin\theta)\varphi'(v)\Big]\beta(\theta)d\theta\\ 
&\ \ \ \ +K_{\beta_\epsilon}^{\varphi}(v,v_{*})-b_\epsilon v\varphi'(v)\\
&\approx \int_{-\epsilon}^\epsilon\Big[v(\cos\theta-1)-v_{*}\sin\theta\Big]^2\beta(\theta)d\theta\frac{\varphi''(v)}{2}+K_{\beta_\epsilon}^{\varphi}(v,v_{*})-b_\epsilon v\varphi'(v)\\
&\approx \frac{\varphi''(v)}{2}v_*^2\int_{-\epsilon}^\epsilon\sin^2\theta\beta(\theta)d\theta+\frac{\varphi''(v)}{2}v^2\int_{-\epsilon}^\epsilon(\cos\theta-1)^2\beta(\theta)d\theta\\
&\ \ \ \ +K_{\beta_\epsilon}^{\varphi}(v,v_{*})-b_\epsilon v\varphi'(v)\\
&\approx \varphi''(v)v_*^2b_\epsilon+K_{\beta_\epsilon}^{\varphi}(v,v_{*})-b_\epsilon v\varphi'(v).
\end{align*}
We decided to neglect the second term in the fourth line of this approximate equality, since it is much smaller than the other terms, because $\int_{-\epsilon}^\epsilon(\cos\theta-1)^2\beta(\theta)d\theta\leq\int_{-\epsilon}^\epsilon\theta^4\beta(\theta)d\theta\leq\epsilon^2\int_{-\epsilon}^\epsilon\theta^2\beta(\theta)d\theta\approx\epsilon^2\int_{-\epsilon}^\epsilon\sin^2\theta\beta(\theta)d\theta$. In order to obtain an equation preserving the kinetic energy, we replaced $\frac{1}{2}\int_{-\epsilon}^\epsilon\sin^2\theta\beta(\theta)d\theta$ by $b_\epsilon$ (both are approximately equal to $\frac{1}{2}\int_{-\epsilon}^\epsilon\theta^2\beta(\theta)d\theta$). Our second main result is the following.

\begin{theo} \label{petits sauts}
Let $f_0\in\mathcal P_4(\mathbb R)$ and let $\beta$ be a cross section satisfying (\ref{ncutoff}). For $\epsilon\in(0,1)$, we consider $(f_t)_{t\geq0}$ and $(f_t^\epsilon)_{t\geq0}$ solutions of (\ref{Kac}) and (\ref{mixte}) respectively, both starting from $f_0$. Then for any $T>0$, any $\epsilon\in(0,1)$, we have 
$$\sup_{t\in[0,T]} W_2^2(f_t,f_t^\epsilon)\leq C\epsilon^2\min\Big(1+T,\frac{1}{\int_{|\theta|<\epsilon}\theta^2\beta(\theta)d\theta}\Big),$$
where $C$ depends only on $\int_{\mathbb R} v^2f_0(dv)$, $\int_{\mathbb R} v^4f_0(dv)$ and on $\int_0^\pi \theta^2\beta(\theta)d\theta$.
\end{theo}
\vskip0.5cm

We can observe that we are not so far to get a bound uniform in time for $\epsilon^2$ (we do not have exponential bounds).

\begin{rmq}
If $\beta$ is as in (\ref{beta}), we get a bound in $C\min\Big(\epsilon^2(1+T),\epsilon^\nu\Big)$.
\end{rmq}

\subsection{System of particles} \label{psystem}

Let $f_0$  be a probability measure on $\mathbb R$ and let $\beta$ be a cross section satisfying (\ref{ncutoff}). We fix an integer $n$ and we consider:
\begin{itemize}
	\item a family of i.i.d. random variables $(V_0^i)_{i\in\{1,...,n\}}$ with law $f_0$,
	\item a family of i.i.d. Poisson measures $(N^{i,n}(ds d\theta dj))_{i\in\{1,...,n\}}$ on $[0,\infty)\times[-\pi,\pi]\times\{1,...,n\}$ with intensity measure $ds\beta(\theta)d\theta\frac{1}{n}\sum_{k=1}^n\delta_k(dj)$,
	\item a family of i.i.d. Brownian motions $(B_t^i)_{t\geq0,\ i\in\{1,...,n\}}$.
\end{itemize}
For $\epsilon\in(0,1)$, we consider $(V_t^{i,n,\epsilon})_{t\geq0,\ i\in\{1,...,n\}}$ solution of the following system of SDEs: for $i=1,...,n$, for all $t\geq0$,
\begin{align} \label{system}
V_t^{i,n,\epsilon}=&V_0^i+\int_0^t\int_{|\theta|>\epsilon}\int_{j\in\{1,...,n\}} \Big[(\cos\theta-1)V_{s-}^{i,n,\epsilon}-\sin\theta V_{s-}^{j,n,\epsilon}\Big]N^{i,n}(dsd\theta dj)\\
& -b_\epsilon\int_0^t V_s^{i,n,\epsilon}ds + \sqrt{2\mathcal E b_\epsilon} B_t^i, \nonumber
\end{align}
where $\mathcal E=\int_{\mathbb R} v^2f_0(dv)$ and $b_\epsilon=\int_{|\theta|<\epsilon}(1-\cos\theta)\beta(\theta)d\theta$.

The quantity $V_t^{i,n,\epsilon}$ has to be thought as the velocity of the $i$-th particle at time $t$. The behavior of $(V_t^{i,n,\epsilon})_{t\geq0}$ is the following: after an exponential time $\tau$ with parameter $\Lambda_\epsilon=\int_{|\theta|\geq\epsilon}\beta(\theta)d\theta$, it collides with another particle labelled $j$ chosen at random and then we set $V_{\tau}^{i,n,\epsilon}=(\cos\Theta) V_{\tau-}^{i,n,\epsilon}-(\sin\Theta) V_{\tau-}^{j,n,\epsilon}$, where $\Theta$ is $\Lambda_\epsilon^{-1}\beta(\theta)\one_{|\theta|\geq\epsilon}d\theta$-distributed. Between two jumps, $V^{i,n,\epsilon}$ behaves like an Ornstein-Uhlenbeck process
\begin{align*}
V_t^{i,n,\epsilon}=V_s^{i,n,\epsilon}-b_\epsilon\int_s^t V_u^{i,n,\epsilon}du + \sqrt{2\mathcal E b_\epsilon} (B_t^i-B_s^i).
\end{align*}
We can solve explicitly this last SDE and we get 
\begin{align*}
V_t^{i,n,\epsilon}=V_s^{i,n,\epsilon}e^{-b_\epsilon (t-s)}+\sqrt{2\mathcal E b_\epsilon} e^{-b_\epsilon (t-s)}\int_s^te^{b_\epsilon u}dB_u^i.
\end{align*}
Hence the strong existence and uniqueness of a solution $(V_t^{i,n,\epsilon})_{t\geq0,\ i\in\{1,...n\}}$ to (\ref{system}) is straightforward.

We can observe that to simulate our system of particles on $[0,T]$, we need to simulate in mean $nT\int_{|\theta|\geq\epsilon}\beta(\theta)d\theta$ jumps. We thus have a cost of simulation of order $nT\int_{|\theta|\geq\epsilon}\beta(\theta)d\theta$. The fact that we can explicitely solve the previous SDE is fundamental in order to have such a cost of simulation. 

\begin{theo} \label{systeme}
Let $f_0\in\mathcal P_4(\mathbb R)$ and let $\beta$ be a cross section satisfying (\ref{ncutoff}). We consider $(f_t)_{t\geq0}$ solution to the Kac equation (\ref{Kac}) starting from $f_0$. For $n\in\mathbb N^*$ and $\epsilon\in(0,1)$, we consider the solution $(V_t^{i,n,\epsilon})_{t\geq0,\ i\in\{1,...n\}}$ to (\ref{system}). We set $\mu_t^{n,\epsilon}=\frac{1}{n}\sum_1^n \delta_{V_t^{i,n,\epsilon}}$.  Then for any $T>0$, any $n\geq2$ and any $\epsilon\in(0,1)$, we have 
\begin{align*}
\sup_{t\in[0,T]}\mathbb E\Big[W_2^2(f_t,\mu_t^{n,\epsilon})\Big]\leq C(1+T)^3\Big(\epsilon^2+\sup_{[0,T]}\mathbb E\Big[W_2^2(f_t,\mu_t^n)\Big]\Big),
\end{align*}
where $C$ depends only on $\int_{\mathbb R} v^2f_0(dv)$, $\int_{\mathbb R} v^4f_0(dv)$ and on $\int_0^\pi \theta^2\beta(\theta)d\theta$, and where for all $t\geq0$, $\mu_t^{n}=\frac{1}{n}\sum_1^n \delta_{V_t^{i}}$, where $(V_t^{i})_{i\in\{1,...n\}}$ is a family of i.i.d. particles with law $f_t$.
\end{theo}
\vskip0.5cm

Applying Lemma \ref{empirique} of the appendix we will deduce the following consequence:
\begin{coro} \label{corosysteme}
Under the same assumptions and notation as in Theorem \ref{systeme}, if $f_0$ has a moment of order $p\geq4$ with $p$ even, then for all $T>0$, all $n\geq2$ and all $\epsilon\in(0,1)$,
\begin{align*}
\sup_{t\in[0,T]}\mathbb E\Big[W_2^2(f_t,\mu_t^{n,\epsilon})\Big]\leq C(1+T)^3\Big(\epsilon^2+\frac{1}{n^{\frac{p-2}{2p-2}}}\Big),
\end{align*}
where $C$ depends only on $p$, $f_0$ and $\int_0^\pi \theta^2\beta(\theta)d\theta$.
\end{coro}

We end this section with a result using another Wasserstein distance.
\begin{prop} \label{W1}
Under the same assumptions and notation as in Theorem \ref{systeme}, if the cross section $\beta$ satisfies the stronger assumption $\int_0^\pi\theta\beta(\theta)d\theta<\infty$, then for all $T>0$, all $n\geq2$ and all $\epsilon\in(0,1)$,
\begin{align*}
\sup_{t\in[0,T]}\mathbb E\Big[W_1(f_t,\mu_t^{n,\epsilon})\Big]\leq C_T\Big(\epsilon+\frac{1}{\sqrt n}\Big),
\end{align*}
where $C_T$ depends only on $T$, $\int_{\mathbb R} v^4f_0(dv)$ and on $\int_0^\pi \theta\beta(\theta)d\theta$.
\end{prop}

We thus have a better dependence in $n$, but we get exponential bounds in time.

\section{Probabilistic interpretation of the equations}

This section is strongly inspired by Tanaka \cite{TAN} and Desvillettes-Graham-M\'el\'eard \cite{MEL}. Until the end of the article,  $(\Omega,\mathcal{F},(\mathcal{F}_t)_{t\geq0},\mathbb P)$ will designate a Polish filtered probability space satisfying the usual conditions. Such a space is Borel isomorphic to the Lebesgue space $([0,1],\mathcal{B}([0,1]),d\alpha)$ which we will use as an auxiliary space. To be as clear as possible, we will use the notation $\mathbb E$ for the expectation and $\mathcal{L}$ for the law of a random variable or process defined on $(\Omega,\mathcal{F},\mathbb P)$, and we will use the notation $\mathbb E_{\alpha}$ and $\mathcal{L}_{\alpha}$ for the expectation and law of random variables or processes on $([0,1],\mathcal{B}([0,1]),d\alpha)$. The processes on $([0,1],\mathcal{B}([0,1]),d\alpha)$ will be called $\alpha$-processes.

We say that a $\mathbb R$-valued process $(V_t)_{t\geq0}$ is a $L^2$-process if it is c\`adl\`ag, adapted and if $\mathbb E(\sup_{[0,T]}V_t^2)<\infty$ for all $T\geq0$.

Now, we introduce a nonlinear stochastic differential equation linked with (\ref{Kac}). 
\begin{prop} \label{def}
Let $\beta$ be a cross section satisfying (\ref{ncutoff}). Let $f_0\in\mathcal P_2(\mathbb R)$ and let $(f_t)_{t\geq0}$ be the solution to (\ref{Kac}) starting from $f_0$. Consider any $\alpha$-process $(W_t)_{t\geq0}$ such that $\mathcal L_\alpha(W_t)=f_t$ for all $t\geq0$. Let also $N$ be a $(\mathcal F_t)_{t\geq0}$-Poisson measure on $[0,\infty)\times [0,1]\times[-\pi,\pi]$ with intensity measure $dsd\alpha\beta(\theta)d\theta$, and $V_0$ a $\mathcal F_0$-measurable random variable with law $f_0$. Then there exists a unique $L^2$-process $(V_t)_{t\geq0}$ such that for all $t\geq0$,
\begin{align} \label{eds}
V_t=V_0+\int_0^t\int_0^1\int_{-\pi}^\pi \Big[(\cos\theta-1)V_{s-}-\sin\theta W_{s-}(\alpha)\Big]\tilde N(dsd\alpha d\theta)-b\int_0^t V_sds,
\end{align}
with $b$ given by (\ref{b}). Furthermore, $\mathcal L(V_t)=f_t$ for all $t\geq0$.
\end{prop}

\textbf{Proof.}
While stated in a slightly different way, this result is almost contained in Desvillettes-Graham-M\'el\'eard \cite[Theorem 3.4]{MEL}. See the proof of Proposition \ref{loi} below for similar arguments. \hfill$\square$
\vskip0.5cm

Let us now write down a probabilistic interpretation of (\ref{EDPmixte}).

\begin{prop} \label{loi2}
Let $g_0\in \mathcal P_2(\mathbb R)$ and set $\mathcal E=\int_{\mathbb R}v^2g_0(dv)$. Consider a $\mathcal F_0$-measurable random variable $Y_0$ with law $g_0$ and a $(\mathcal F_t)_{t\geq0}$-Brownian motion $(B_t)_{t\geq0}$. Then there exists a unique $L^2$-process $(Y_t)_{t\geq0}$ such that for all $t\geq0$,
\begin{align} \label{Y}
 Y_t=Y_0-\frac{1}{2}\int_0^t Y_sds + \sqrt{\mathcal{E}}B_t.
\end{align}
Furthermore, $\mathcal L(Y_t)=g_t$ for all $t\geq0$, where $(g_t)_{t\geq0}$ is the unique solution to (\ref{EDPmixte}).
\end{prop}

\textbf{Proof.} The existence and uniqueness of $Y$ is classical since (\ref{Y}) is a S.D.E. with Lipschitz coefficients. By It\^o's formula, we have for any $\varphi\in C_b^2(\mathbb R)$
\begin{align*}
\varphi(Y_t)=\varphi(Y_0)+\int_0^t\varphi'(Y_s)(-\frac{1}{2}Y_sds+\sqrt{\mathcal E}dB_s)+\frac{\mathcal E}{2}\int_0^t\varphi''(Y_s)ds.
\end{align*}
Taking expectations and setting $\mu_t=\mathcal L(Y_t)$, we get for any $\varphi\in C_b^2(\mathbb R)$
\begin{align*}
\int_{\mathbb R} \varphi(v)\mu_t(dv)=\int_{\mathbb R} \varphi(v)g_0(dv)-\frac{1}{2}\int_0^t\int_{\mathbb R} v\varphi'(v)\mu_s(dv)ds+\frac{\mathcal E}{2}\int_0^t\int_{\mathbb R} \varphi''(v)\mu_s(dv).
\end{align*}
Thus $(\mu_t)_{t\geq0}$ solves (\ref{EDPmixte}) in the sense of Definition \ref{weak solutions}. We get $(\mu_t)_{t\geq0}=(g_t)_{t\geq0}$ by uniqueness (see Proposition \ref{ex}). \hfill$\square$
\vskip0.5cm

It remains to give a probabilistic interpretation of (\ref{mixte}).
\begin{prop} \label{loi}
Let $\epsilon\in(0,1)$ be fixed. Consider a cross-section $\beta$ satisfying (\ref{ncutoff}), a probability measure $f_0^\epsilon\in\mathcal P_2(\mathbb R)$, and the corresponding unique solution $(f_t^\epsilon)_{t\geq0}$ to (\ref{mixte}). Consider any $\alpha$-process $(W_t^\epsilon)_{t\geq0}$ such that for all $t\geq0$, $\mathcal L_\alpha(W_t^\epsilon)=f_t^\epsilon$. Let $V_0^\epsilon$ be a $\mathcal F_0$-measurable random variable with law $f_0^\epsilon$, let $N$ be a $(\mathcal F_t)_{t\geq0}$-Poisson measure on $[0,\infty)\times [0,1]\times[-\pi,\pi]$ with intensity measure $dsd\alpha\beta(\theta)d\theta$ and let $(B_t)_{t\geq0}$ be a $(\mathcal F_t)_{t\geq0}$-Brownian motion independent of $N$. Then there exists a unique $L^2$-process $(V_t^\epsilon)_{t\geq0}$ such that for all $t\geq0$,
\begin{align}
\label{Veps}
V_t^\epsilon=&V_0^\epsilon+\int_0^t\int_{|\theta|\geq\epsilon}\int_0^1 \Big[(\cos\theta-1)V_{s-}^\epsilon-\sin\theta W_{s-}^\epsilon(\alpha)\Big] N(d\theta d\alpha ds)\\
\nonumber
&-b_\epsilon\int_0^t V_s^\epsilon ds + \sqrt{2\mathcal E b_\epsilon} B_t, 
\end{align}
with $b_\epsilon$ defined in (\ref{beps2}). Furthermore, $ \mathcal L(V_t^\epsilon)=f_t^\epsilon$ for all $t\geq0$.
\end{prop}

\textbf{Proof.} See Ikeda-Watanabe \cite[Theorem 9.1]{IKE} for existence and uniqueness of $(V_t^\epsilon)_{t\geq0}$: (\ref{Veps}) is a classical jumping S.D.E. with Lipschitz coefficients. Let $\varphi\in C_b^2(\mathbb R)$. By It\^o's formula for jump processes (see e.g. Ikeda-Watanabe \cite[Theorem 5.1]{IKE}), we have
\begin{align*}
\varphi(V_t^\epsilon)=&\varphi(V_0^\epsilon)+\int_0^t\varphi'(V_s^\epsilon)(-b_\epsilon V_s^\epsilon ds+\sqrt{2\mathcal E b_\epsilon}dB_s)+\mathcal E b_\epsilon\int_0^t\varphi''(V_s^\epsilon)ds\\
&+\int_0^t\int_{|\theta|\geq\epsilon}\int_0^1 \Big[\varphi(\cos\theta V_{s-}^\epsilon-\sin\theta W_{s-}^\epsilon(\alpha))-\varphi(V_{s-}^\epsilon)\Big] N(d\theta d\alpha ds).
\end{align*}
Taking expectations and setting $\mu_t^\epsilon=\mathcal L(V_t^\epsilon)$, we get for any $\varphi\in C_b^2(\mathbb R)$, using that $\mathcal L_\alpha(W_t^\epsilon)=f_t^\epsilon$,
\begin{align*}
\int_{\mathbb R} \varphi(v)&\mu_t^\epsilon(dv)=\int_{\mathbb R} \varphi(v)f_0^\epsilon(dv)-b_\epsilon\int_0^t\int_{\mathbb R} v\varphi'(v)\mu_s^\epsilon(dv)ds\\
&+\mathcal E b_\epsilon\int_0^t\int_{\mathbb R} \varphi''(v)\mu_s^\epsilon(dv)ds\\
&+\int_0^t\int_{\mathbb R}\int_{\mathbb R}\int_{|\theta|\geq\epsilon} \Big[\varphi(v\cos\theta-v_*\sin\theta)-\varphi(v)\Big]\beta(\theta)d\theta\mu_s^\epsilon(dv)f_s^\epsilon(dv_*)ds.
\end{align*}
But $(f_t^\epsilon)_{t\geq0}$ solves the same equation since it solves (\ref{mixte}) in the sense of Definition \ref{weak solutions}. Since $(f_s^\epsilon)_{s\geq0}$ is given, this equation is linear and we have uniqueness of the solution. Indeed, we use Proposition \ref{unicite} with $a=\mathcal E b_\epsilon$, $b=-b_\epsilon$, $r=0$ and $q(t,v,A)=\int_{-\pi}^\pi\int_{\mathbb R} \one_{|\theta|>\epsilon}\one_A\big(v(\cos\theta-1)-v_*\sin\theta \big)f_t^\epsilon(dv_*)\beta(\theta)d\theta$ for all Borel subset $A\subset\mathbb R$, which satisfies $\sup_{t,v}q(t,v,\mathbb R)=\int_{|\theta|>\epsilon}\beta(\theta)d\theta<\infty$ and $\sup_{t\geq0}\int_{\mathbb R} (h^2+2vh)q(t,v,dh)=\int_{|\theta|>\epsilon}\sin^2\theta\beta(\theta)d\theta\Big(\int_{\mathbb R}v_*^2f_0(dv_*)+v^2\Big)\leq C(1+v^2)$. Finally, we get $(\mu_t^\epsilon)_{t\geq0}=(f_t^\epsilon)_{t\geq0}$. \hfill$\square$

\section{The Grazing collisions limit}

We consider a family of cross sections $(\beta_\epsilon)_{\epsilon\in(0,1)}$ with $\int_{-\pi}^{\pi}\theta^2 \beta_{\epsilon}(\theta)d\theta=1$ and $\int_{-\pi}^{\pi}\theta^4 \beta_{\epsilon}(\theta)d\theta\stackrel{\epsilon\rightarrow0}{\longrightarrow}0$. Let $g_0\in\mathcal P_4(\mathbb R)$. For any $\epsilon\in(0,1)$, we consider $(g_t^\epsilon)_{t\geq0}$ the unique solution of (\ref{Kac}) with cross section $\beta_\epsilon$ starting from $g_0$. We also consider $(g_t)_{t\geq0}$ the unique solution of (\ref{EDPmixte}) starting from $g_0$. For $\epsilon\in(0,1)$, we consider a $\mathcal F_0$-measurable random variable $V_0$ with law $g_0$, and a $(\mathcal F_t)_{t\geq0}$-Poisson measure $N^\epsilon$ on $[0,\infty)\times [0,1]\times[-\pi,\pi]$ with intensity measure $dsd\alpha\beta_\epsilon(\theta)d\theta$. We also consider an $\alpha$-process $(W_t^\epsilon)_{t\geq0}$ such that $\mathcal L_\alpha(W_t^\epsilon)=g_t^\epsilon$ for all $t\geq0$. Let $(B_t)_{t\geq0}$ be a $(\mathcal F_t)_{t\geq0}$-Brownian motion. We consider $(V_t^{\epsilon})_{t\geq0}$ and $(Y_t)_{t\geq0}$ solutions of the following S.D.E.s
\begin{align*} 
V_t^\epsilon=V_0+\int_0^t\int_0^1\int_{-\pi}^\pi \Big[(\cos\theta-1)V_{s-}^\epsilon-\sin\theta W_{s-}^\epsilon(\alpha)\Big]\tilde N^\epsilon(dsd\alpha d\theta)-b_\epsilon\int_0^t V_s^\epsilon ds,
\end{align*}
where $b_\epsilon=\int_{-\pi}^\pi(1-\cos\theta)\beta_\epsilon(\theta)d\theta$ and
\begin{align} \label{Y2}
Y_t=V_0-\frac{1}{2}\int_0^t Y_sds + \sqrt{\mathcal{E}}B_t.
\end{align}
 
Theorem \ref{collisions rasantes} is a corollary of the following statement.

\begin{theo}  \label{conclusion}
For any $t\geq0$ and any $\epsilon\in(0,1)$, we can couple the Poisson measure $N^\epsilon$ and the Brownian motion $B$ in such a way that 
\begin{align*}  
\mathbb E[(V_t^{\epsilon}-Y_t)^2]\leq & 4\Big[\frac{\mathcal{E}\int_{-\pi}^{\pi}(1-\cos\theta)^2\beta_{\epsilon}(\theta)d\theta}{2b_{\epsilon}}+C\frac{(\mathbb E(V_0^4)+ 3\mathcal{E}^2)\int_{-\pi}^{\pi} \sin^4\theta \beta_{\epsilon}(\theta)d\theta }{\mathcal{E}\gamma_{\epsilon}}\\
&+\mathcal{E}\Big(|\ln(2b_\epsilon)|^2+\Big(\frac{\gamma_\epsilon}{2b_\epsilon}+1\Big)|2b_\epsilon-1|+2|\gamma_\epsilon-1|\Big)\Big],
\end{align*}
where $b_{\epsilon}=\int_{-\pi}^{\pi}(1-\cos\theta)\beta_{\epsilon}(\theta)d\theta$, $\gamma_{\epsilon}=\int_{-\pi}^{\pi} \sin^2\theta \beta_{\epsilon}(\theta)d\theta$, $\mathcal E=\mathbb E[V_0^2]$ and $C$ is a universal constant.
\end{theo}
Let us insist on the fact that the coupling between $N^\epsilon$ and $B$ depends on $t$.
Assuming for a moment that this result holds true, we can prove Theorem \ref{collisions rasantes}.
\vskip0.5cm

\textbf{Proof of Theorem \ref{collisions rasantes}.}
First recalling that $\mathcal L(V_t^\epsilon)=g_t^\epsilon$ by Proposition \ref{def} and $\mathcal L(Y_t)=g_t$ by Proposition \ref{loi2}, we have $W_2^2(g_t^\epsilon,g_t)\leq \mathbb E[(V_t^{\epsilon}-Y_t)^2]$. If $\int_{-\pi}^{\pi}\theta^4\beta_\epsilon(\theta)d\theta>1$, we have $\mathbb E[(V_t^{\epsilon}-Y_t)^2]\leq 2\mathbb E[(V_t^\epsilon)^2]+2\mathbb E[Y_t^2]=4\mathcal E\leq 4\mathcal E\int_{-\pi}^{\pi}\theta^4\beta_\epsilon(\theta)d\theta$. We now suppose that $\int_{-\pi}^{\pi}\theta^4\beta_\epsilon(\theta)d\theta<1$. Using the Taylor-Lagrange inequality, we have
\begin{align*}
|1-&\cos\theta|\leq \theta^2/2,\quad |2(1-\cos\theta)-\theta^2|\leq\theta^4/12\\
&|\sin \theta|\leq|\theta|\quad \text{and}\quad |\sin^2\theta-\theta^2|\leq\theta^4/3.
\end{align*}
Using these inequalities, we get
\begin{align*}
\int_{-\pi}^{\pi}(1-\cos\theta)^2\beta_{\epsilon}(\theta)d\theta&\leq\int_{-\pi}^{\pi}(\theta^4/4)\beta_{\epsilon}(\theta)d\theta,\quad \int_{-\pi}^{\pi} \sin^4\theta \beta_{\epsilon}(\theta)d\theta\leq\int_{-\pi}^{\pi}\theta^4\beta_{\epsilon}(\theta)d\theta,
\end{align*} 
and, recalling that $\int_{-\pi}^{\pi}\theta^2 \beta_{\epsilon}(\theta)d\theta=1$, 
\begin{align*}
|2b_\epsilon-1|\leq\int_{-\pi}^{\pi}(\theta^4/12)\beta_{\epsilon}(\theta)d\theta,\quad |\gamma_\epsilon-1|\leq \int_{-\pi}^{\pi}(\theta^4/3)\beta_{\epsilon}(\theta)d\theta.
\end{align*}
Since $\int_{-\pi}^{\pi}\theta^4 \beta_{\epsilon}(\theta)d\theta<1$ by assumption, we have $2b_\epsilon\geq11/12$, $2/3\leq\gamma_\epsilon\leq4/3$ and $2b_{\epsilon}-1\in[-1/12,1/12]$ which allows us to write $|\ln(2b_{\epsilon})|^2=|\ln(1+(2b_{\epsilon}-1))|^2\leq 4|2b_{\epsilon}-1|^2$. We thus get $\mathbb E[(V_t^{\epsilon}-Y_t)^2]\leq C(\mathcal E+\frac{\mathbb E(V_0^4)}{\mathcal E})\int_{-\pi}^{\pi}\theta^4\beta_{\epsilon}(\theta)d\theta$, which concludes the proof, since $\mathcal E^2=\mathbb E[V_0^2]^2\leq\mathbb E[V_0^4]$ by the Cauchy-Schwarz inequality. \hfill$\square$ 
\vskip1cm

It remains to prove Theorem \ref{conclusion}. Let us start with the following lemma.
\begin{lemme} \label{lemme1}
For $\epsilon\in(0,1)$, let $Y^{\epsilon}$ be the unique solution of
\begin{align} 
Y_t^{\epsilon}=V_0 - b_{\epsilon}\int_0^t Y_s^{\epsilon}ds - \int_0^t\int_0^1\int_{-\pi}^\pi \sin\theta W_{s-}^{\epsilon}(\alpha) \tilde N^{\epsilon}(dsd\alpha d\theta)
\end{align}
(since $W^\epsilon$ is a given $\alpha$-process, this is a classical S.D.E. with Lipschitz coefficients). Then for all $t\geq0$, 
$$\mathbb E\Big((V_t^{\epsilon}-Y_t^{\epsilon})^2\Big)\leq \frac{\mathcal{E}\int_{-\pi}^{\pi}(1-\cos\theta)^2\beta_{\epsilon}(\theta)d\theta}{2b_{\epsilon}}. $$
\end{lemme}

\textbf{Proof.}
Observing that
$$V_t^{\epsilon}-Y_t^{\epsilon}=\int_0^t\int_0^1\int_{-\pi}^\pi(\cos\theta-1)V_{s-}^{\epsilon}\tilde N^{\epsilon}(ds d\alpha d\theta) -b_{\epsilon}\int_0^t (V_s^{\epsilon}-Y_s^{\epsilon})ds,$$
we get by It\^o's formula
\begin{align*}
(V_t^{\epsilon}-Y_t^{\epsilon})^2 &= \int_0^t\int_0^1\int_{-\pi}^{\pi} \Big[\Big(V_{s-}^{\epsilon}-Y_{s-}^{\epsilon}+(\cos\theta-1)V_{s-}^{\epsilon}\Big)^2\\
&\ \ \ \ \ \ \ \ \ \ \ \ \ \ \ \ \ \ \ \ -(V_{s-}^{\epsilon}-Y_{s-}^{\epsilon})^2\Big]\tilde N^{\epsilon}(ds d\alpha d\theta)\\
&\ \ \ + 	\int_0^t\int_0^1\int_{-\pi}^\pi \Big[\Big(V_{s}^{\epsilon}-Y_{s}^{\epsilon}+(\cos\theta-1)V_{s}^{\epsilon}\Big)^2-(V_{s}^{\epsilon}-Y_{s}^{\epsilon})^2\\
&\ \ \ \quad\quad\quad\quad\quad\quad\quad-2(V_s^{\epsilon}-Y_s^{\epsilon})(\cos\theta-1)V_{s}^{\epsilon}\Big]dsd\alpha\beta_{\epsilon}(\theta)d\theta\\
&\ \ \ -2b_{\epsilon}\int_0^t (V_s^{\epsilon}-Y_s^{\epsilon})^2ds\\
&=M_t^\epsilon+\int_0^t\int_{-\pi}^{\pi} (\cos\theta-1)^2(V_s^{\epsilon})^2ds\beta_{\epsilon}(\theta)d\theta -2b_{\epsilon}\int_0^t(V_s^{\epsilon}-Y_s^{\epsilon})^2ds,
\end{align*}
where $M_t^\epsilon$ is a martingale with mean 0. So using that $\mathbb E[(V_t^\epsilon)^2]=\int_{\mathbb R} v^2g_t^\epsilon(dv)=\int_{\mathbb R} v^2g_0(dv)=\mathcal E$ for all $t\geq0$ by (\ref{energie}), we have  
\begin{align*}
\mathbb E[(V_t^{\epsilon}-Y_t^{\epsilon})^2]&=\int_0^t\int_{-\pi}^{\pi}(1-\cos\theta)^2\mathbb E[(V_s^{\epsilon})^2]\beta_{\epsilon}(\theta)d\theta ds-2b_{\epsilon}\int_0^t\mathbb E[(V_s^{\epsilon}-Y_s^{\epsilon})^2]ds\\
&=\mathcal{E}t\int_{-\pi}^{\pi}(1-\cos\theta)^2\beta_{\epsilon}(\theta)d\theta  -2b_{\epsilon}\int_0^t\mathbb E[(V_s^{\epsilon}-Y_s^{\epsilon})^2]ds.
\end{align*}
Differentiating this equality with respect to $t$, we find an O.D.E. that can be solved explicitly. This gives
\begin{align*}
\mathbb E[(V_t^{\epsilon}-Y_t^{\epsilon})^2]=\frac{\mathcal{E}\int_{-\pi}^{\pi}(1-\cos\theta)^2\beta_{\epsilon}(\theta)d\theta}{2b_{\epsilon}}(1-e^{-2b_\epsilon t}).
\end{align*}
The conclusion follows. \hfill$\square$ 
\vskip0.5cm

In the following lemma, using Corollary \ref{coro}, we will find a suitable coupling between our Poisson measure $N^\epsilon$ and our Brownian motion $B$.
\begin{lemme} \label{lemme2}
Let $\tilde Y^{\epsilon}$ be the unique solution of 
\begin{align} \label{Ytilde}
\tilde Y_t^{\epsilon}=V_0-b_\epsilon\int_0^t\tilde Y_s^{\epsilon}ds+\sqrt{\mathcal{E}\gamma_{\epsilon}}B_t.
\end{align}
We consider the process $Y^\epsilon$ defined in Lemma \ref{lemme1}. For any $\epsilon\in(0,1)$ and for each $t\geq0$, we can couple the Poisson measure $N^\epsilon$ and the Brownian motion $B$ in such a way that
$$\mathbb E[(\tilde Y_t^{\epsilon}-Y_t^{\epsilon})^2]\leq C\frac{(\mathbb E(V_0^4)+ 3\mathcal{E}^2)\int_{-\pi}^{\pi} \sin^4\theta \beta_{\epsilon}(\theta)d\theta }{\mathcal{E}\gamma_{\epsilon}},$$
where $C$ is a universal constant and $\gamma_\epsilon=\int_{-\pi}^\pi\sin^2\theta\beta_\epsilon(\theta)d\theta$.
\end{lemme}
Observe that for each $t$ we need a suitable coupling. We are not able to find a coupling working simultaneously for all values of $t$. 
\vskip0.5cm

\textbf{Proof.}
Applying It\^o's formula, we get $\tilde Y_t^{\epsilon}e^{b_{\epsilon}t}=V_0+\sqrt{\mathcal{E}\gamma_{\epsilon}}\int_0^te^{b_{\epsilon}s}dB_s$ and $Y_t^{\epsilon}e^{b_{\epsilon}t}=V_0-\int_0^t\int_0^1\int_{-\pi}^\pi e^{b_{\epsilon}s}\sin\theta W_{s-}^{\epsilon}(\alpha)\tilde N^\epsilon(dsd\alpha d\theta)$. We observe that the random variable $\sqrt{\mathcal{E}\gamma_{\epsilon}}\int_0^te^{b_{\epsilon}s}dB_s$ follows a centered normal law with variance $\mathcal{E}\gamma_{\epsilon}\int_0^te^{2b_{\epsilon}s}$ which is equal to $\int_0^t\int_0^1\int_{-\pi}^\pi e^{2b_{\epsilon}s}\sin^2\theta (W_{s}^{\epsilon}(\alpha))^2 \beta_{\epsilon}(\theta)d\theta d\alpha ds$ because $\mathcal L_\alpha(W_s^\epsilon)=g_s^\epsilon$ and due to (\ref{energie}). So using Corollary \ref{coro}, we get
\begin{align*}
W_2^2(\mathcal L(\tilde Y_t^{\epsilon}e^{b_{\epsilon}t}),\mathcal L(Y_t^{\epsilon}e^{b_{\epsilon}t}))&\leq C_0\frac{\int_0^t\int_0^1\int_{-\pi}^\pi e^{4b_{\epsilon}s}\sin^4\theta (W_{s}^{\epsilon}(\alpha))^4 \beta_{\epsilon}(\theta)dsd\alpha d\theta}{\mathcal{E}\gamma_{\epsilon}\int_0^te^{2b_{\epsilon}s}ds}\\
&=C_0\frac{\int_0^t e^{4b_{\epsilon}s}\mathbb E_\alpha((W_{s}^{\epsilon})^4)ds\int_{-\pi}^{\pi} \sin^4\theta \beta_{\epsilon}(\theta)d\theta}{\mathcal{E}\gamma_{\epsilon}\int_0^te^{2b_{\epsilon}s}ds}.
\end{align*}
Using Lemma \ref{moment4}, since $\mathcal L(W_s^\epsilon)=g_s^\epsilon$ and since $g^\epsilon$ solves (\ref{Kac}) (with the cross section $\beta_\epsilon$), we deduce $\mathbb E_\alpha[(W_s^\epsilon)^2]\leq\int_{\mathbb R} v^4g_0(dv)+3\mathcal E^2=\mathbb E[V_0^4]+3\mathcal E^2$ for all $s\geq0$. Hence, using that $\frac{\int_0^te^{4b_\epsilon s}ds}{\int_0^te^{2b_\epsilon s}ds}\leq e^{2b_\epsilon t}$, we have
\begin{align*}
W_2^2(\mathcal L(\tilde Y_t^{\epsilon}e^{b_{\epsilon}t}),\mathcal L(Y_t^{\epsilon}e^{b_{\epsilon}t}))\leq C_0\frac{(\mathbb E(V_0^4)+ 3\mathcal{E}^2)\int_{-\pi}^{\pi} \sin^4\theta \beta_{\epsilon}(\theta)d\theta e^{2b_{\epsilon}t}}{\mathcal{E}\gamma_{\epsilon}}.
\end{align*}
Consequently,
$$W_2^2(\mathcal L(\tilde Y_t^{\epsilon}),\mathcal L(Y_t^{\epsilon}))\leq C_0\frac{(\mathbb E(V_0^4)+ 3\mathcal{E}^2)\int_{-\pi}^{\pi} \sin^4\theta \beta_{\epsilon}(\theta)d\theta }{\mathcal{E}\gamma_{\epsilon}}.$$
To conclude, it suffices to take $N^\epsilon$ and $B$ in such a way that $\mathbb E[(\tilde Y_t^{\epsilon}-Y_t^{\epsilon})^2]=W_2^2(\mathcal L(\tilde Y_t^{\epsilon}),\mathcal L(Y_t^{\epsilon}))$. \hfill$\square$ 
\vskip0.5cm

Let us now give the last lemma needed to prove Theorem \ref{conclusion}.
\begin{lemme} \label{lemme3}
Consider the unique solutions $Y$ and $\tilde Y^\epsilon$ to (\ref{Y2}) and (\ref{Ytilde}) respectively, driven by the same Brownian motion $B$. Then for all $t\geq0$ fixed and for all $\epsilon\in(0,1)$,
$$\mathbb E[(\tilde Y_t^{\epsilon}-Y_t)^2]\leq \mathcal{E}\Big(|\ln(2b_\epsilon)|^2+\Big(\frac{\gamma_\epsilon}{2b_\epsilon}+1\Big)|2b_\epsilon-1|+2|\gamma_\epsilon-1|\Big).$$ 
\end{lemme}

\textbf{Proof.}
We have $\tilde Y_t^{\epsilon}=V_0e^{-b_{\epsilon}t}+\sqrt{\mathcal{E}\gamma_{\epsilon}}e^{-b_{\epsilon}t}\int_0^te^{b_{\epsilon}s}dB_s$ and $Y_t=V_0e^{-t/2}+\sqrt{\mathcal{E}}e^{-t/2}\int_0^te^{s/2}dB_s$ as in the proof of Lemma \ref{lemme2}. Since $B$ and $V_0$ are independent, we have
\begin{align*}
\mathbb E[(\tilde Y_t^{\epsilon}-Y_t)^2]&= \mathbb E(V_0^2)(e^{-b_{\epsilon}t}-e^{-t/2})^2\\
&\ \ \ \ \ +\mathbb E\Big[\Big(\int_0^t(\sqrt{\mathcal{E}\gamma_{\epsilon}}e^{-b_{\epsilon}t}e^{b_{\epsilon}s}-\sqrt{\mathcal{E}}e^{-t/2}e^{s/2})dB_s\Big)^2\Big]\\
&=\mathcal{E}(e^{-b_{\epsilon}t}-e^{-t/2})^2+ \mathcal E\int_0^t(\sqrt{\gamma_{\epsilon}}e^{-b_{\epsilon}(t-s)}-e^{-(t-s)/2})^2ds\\
&=\mathcal{E}(e^{-b_{\epsilon}t}-e^{-t/2})^2+ \mathcal E\int_0^t(\sqrt{\gamma_{\epsilon}}e^{-b_{\epsilon}s}-e^{-s/2})^2ds.
\end{align*}
We set $h(t)=(e^{-b_{\epsilon}t}-e^{-t/2})^2$. The function $h$ reaches its maximum at $t_0=\ln(2b_{\epsilon})/(b_{\epsilon}-1/2)$. Moreover, $|h(t_0)|=|e^{-b_{\epsilon}t_0}-e^{-t_0/2}|^2\leq |b_\epsilon t_0-t_0/2|^2=|t_0|^2|b_{\epsilon}-1/2|^2=|\ln(2b_{\epsilon})|^2$. Next,

\begin{align*}
\int_0^t(\sqrt{\gamma_{\epsilon}}e^{-b_{\epsilon}s}-e^{-s/2})^2ds&\leq\int_0^\infty(\sqrt{\gamma_{\epsilon}}e^{-b_{\epsilon}s}-e^{-s/2})^2ds\\
&=\frac{\gamma_\epsilon}{2b_\epsilon}+1-\frac{2\sqrt{\gamma_\epsilon}}{b_\epsilon+1/2}\\
&=\frac{\gamma_\epsilon(b_\epsilon+1/2)+2b_\epsilon(b_\epsilon+1/2)-4\sqrt{\gamma_\epsilon}b_\epsilon}{2b_\epsilon(b_\epsilon+1/2)}\\
&\leq\frac{1}{b_\epsilon}\Big[\gamma_\epsilon(1/2-b_\epsilon)+2b_\epsilon(\gamma_\epsilon+b_\epsilon+1/2-2\sqrt{\gamma_\epsilon})\Big]\\
&=\frac{\gamma_\epsilon}{2b_\epsilon}(1-2b_\epsilon)+2\Big[(\sqrt{\gamma_\epsilon}-1)^2+(b_\epsilon-1/2)\Big]\\
&\leq\Big(\frac{\gamma_\epsilon}{2b_\epsilon}+1\Big)|2b_\epsilon-1|+2|\gamma_\epsilon-1|,
\end{align*}
the last inequality coming from $(\sqrt x-\sqrt y)^2\leq|x-y|$. The lemma is proved. \hfill$\square$

\vskip1cm
We can now conclude this section.
\vskip0.5cm

\textbf{Proof of Theorem \ref{conclusion}.}
For $\epsilon\in(0,1)$ and $t\geq0$ fixed, we take the Poisson measure $N^\epsilon$ and the Brownian motion $B$ as in Lemma \ref{lemme2} and we consider the processes $V^\epsilon$, $Y$, $Y^\epsilon$ and $\tilde Y^\epsilon$ build with this $N^\epsilon$ and this $B$. Then, writing
\begin{align*}
\mathbb E[(V_t^\epsilon-Y_t)^2]&\leq 4\Big[\mathbb E[(V_t^\epsilon-Y_t^\epsilon)^2]+\mathbb E[(Y_t^\epsilon-\tilde Y_t^\epsilon)^2]+\mathbb E[(\tilde Y_t^\epsilon-Y_t)^2]\Big],
\end{align*}
and using Lemmas \ref{lemme1}, \ref{lemme2} and \ref{lemme3}, we immediately conclude. \hfill$\square$

\section{Cutoff approximation with diffusion} 

The whole section is dedicated to the proof of Theorem \ref{petits sauts}. Let thus $f_0\in\mathcal P_4(\mathbb R)$ and let $\beta$ be a cross section satisfying (\ref{ncutoff}). We fix $\epsilon\in(0,1)$, and we consider the solutions $(f_t)_{t\geq0}$ and $(f_t^\epsilon)_{t\geq0}$ to (\ref{Kac}) and (\ref{mixte}) respectively, both starting from $f_0$.
\vskip0.5cm

We will proceed as follows. We fix some $t_0\geq0$ for the whole proof. We will build some solutions $(V_t)_{t\geq0}$ and $(V_t^\epsilon)_{t\geq0}$ to (\ref{eds}) and (\ref{Veps}), both starting from some initial value $V_0$ with law $f_0$, coupled in such a way that $\mathbb E[(V_{t_0}-V_{t_0}^\epsilon)^2]$ is as small as possible.
\vskip0.5cm

We divide the proof into five steps. In the first step, we  introduce the (suitably coupled) processes $(V_t)_{t\geq0}$, $(V_t^\epsilon)_{t\geq0}$ as well as an intermediate process $(\tilde V_t^\epsilon)_{t\geq0}$. In Step 2, we upperbound $\mathbb E[(V_{t_0}^\epsilon-\tilde V_{t_0}^\epsilon)^2]$. Step 3 is dedicated to the study of $\mathbb E[(\tilde V_{t_0}^\epsilon-V_{t_0})^2]$. In Step 4, we show that $\mathbb E[(V_{t_0}^\epsilon-\tilde V_{t_0}^\epsilon)(\tilde V_{t_0}^\epsilon-V_{t_0})]=0$. We conclude in Step 5.
\vskip0.5cm
In the whole section, we will use the notation
\begin{align} \label{ceps}
&b_\epsilon=\int_{|\theta|<\epsilon}(1-\cos\theta)\beta(\theta)d\theta, \quad c_\epsilon=2b_\epsilon+\int_{|\theta|\geq\epsilon}\sin^2\theta\beta(\theta)d\theta, \\
\nonumber
&d_\epsilon=\int_{|\theta|\geq\epsilon}\sin^2\theta\beta(\theta)d\theta \quad \text{and} \quad \gamma_\epsilon=\int_{|\theta|<\epsilon}(1-\cos\theta)^2\beta(\theta)d\theta.
\end{align}

\underline{Step 1}: the coupling.

- Let $(\Omega_i,\mathcal{F}^i, (\mathcal{F}_t^i)_{t\geq0}, \mathbb{P}_i)$, $i=1,2$, be two Polish filtered probability spaces satisfying the usual conditions and consider the following filtered probability space $(\Omega,\mathcal{F},(\mathcal{F}_t)_{t\geq0},\mathbb{P})=(\Omega_1\times\Omega_2, \mathcal{F}^1\otimes\mathcal{F}^2, (\mathcal{F}_t^1\otimes\mathcal{F}_t^2)_{t\geq0}, \mathbb{P}_1\otimes\mathbb{P}_2)$. We denote by $\mathbb{E}$ the expectation under $\mathbb{P}$ and by $\mathbb{E}_i$ the expectation under $\mathbb{P}_i$.  

- On $(\Omega_1,\mathcal{F}^1, (\mathcal{F}_t^1)_{t\geq0}, \mathbb{P}_1)$, we consider a $f_0$-distributed random variable $V_0$ $\mathcal F_0^1$-measurable, as well as a $(\mathcal F_t^1)_{t\geq0}$-Poisson measure $N_{|\theta|\geq\epsilon}$ on $[0,\infty)\times[0,1]\times[-\pi,\pi]$ with intensity measure $dsd\alpha\beta(\theta)\one_{|\theta|\geq\epsilon}d\theta$. We set 
\begin{align} \label{X}
X_t=\int_0^t\int_0^1\int_{-\pi}^\pi (\cos\theta-1)N_{|\theta|\geq\epsilon}(dsd\alpha d\theta) -b_{\epsilon}t,
\end{align}
We consider the Dol\'eans-Dade exponential of $X$, see Jacod-Shiryaev \cite[Theorem 4.61]{JAC}, defined by
\begin{align} 
Z_t=1+\int_0^t Z_{s-}dX_s.
\end{align}
There holds 
\begin{align} \label{doleans}
Z_t=e^{X_t}\prod_{s\leq t}(1+\Delta X_s)e^{-\Delta X_s}=e^{-b_\epsilon t}\prod_{i\geq 1}\cos\theta_i\one_{T_i\leq t},
\end{align}
where $(T_i,\theta_i,\alpha_i)_{i\geq1}$ are the marks of the Poisson measure $N_{|\theta|\geq\epsilon}$. Observe that a.s., $Z_t\neq0$ $\forall t\geq0$, because $\beta(\theta)d\theta$ does not give weight to $\Big\{-\frac{\pi}{2},\frac{\pi}{2}\Big\}$.

Of course, the processes $(X_t)_{t\geq0}$ and $(Z_t)_{t\geq0}$ depend on $\epsilon$ but we do not write this dependence in order to lighten notations.

- For  each $t\geq 0$, we consider some $\alpha$-random variables $W_t$ and $W_t^\epsilon$ with respective laws $f_t$ and $f_t^\epsilon$ verifying
\begin{align} \label{delta}
W_2^2(f_t,f_t^\epsilon)=\mathbb E_\alpha[(W_t-W_t^\epsilon)^2].
\end{align}

- Recall that $t_0\geq0$ is fixed. Fix also $\omega_1\in\Omega_1$. On $(\Omega_2,\mathcal{F}^2, (\mathcal{F}_t^2)_{t\geq0}, \mathbb P_2)$, we consider a $(\mathcal F_t^2)_{t\geq0}$-Poisson measure $N_{|\theta|<\epsilon}^{\omega_1}$ on $[0,\infty)\times[0,1]\times[-\pi,\pi]$ with intensity measure $dsd\alpha\beta(\theta)\one_{|\theta|<\epsilon}d\theta$ and a Brownian motion $(B_t^{\omega_1})_{t\geq0}$ (we do not write the dependence in $t_0$ and $\epsilon$) such that:
\begin{align}
\nonumber
W_2^2(\mu_{t_0}^{\omega_1},\nu_{t_0}^{\omega_1})=\mathbb{E}_2\Big[\Big(&\int_0^{t_0}\int_0^1\int_{-\pi}^\pi (Z_{t_0}Z_{s-}^{-1})(\omega_1)\sin(-\theta)W_{s-}(\alpha)\tilde N_{|\theta|<\epsilon}^{\omega_1}(\omega_2,dsd\alpha d\theta)\\
\label{wasser}
&-\int_0^{t_0} \sqrt{2\mathcal E b_\epsilon} (Z_{t_0}Z_s^{-1})(\omega_1)dB_s^{\omega_1}(\omega_2)\Big)^2\Big],
\end{align}
where
\begin{align} \label{mut}
\mu_{t_0}^{\omega_1}=\mathcal{L}_2\Big(\int_0^{t_0}\int_0^1\int _{-\pi}^\pi(Z_{t_0}Z_{s-}^{-1})(\omega_1)\sin(-\theta)W_{s-}(\alpha)\tilde N_{|\theta|<\epsilon}^{\omega_1}(\omega_2,ds d\alpha d\theta)\Big),
\end{align}
\begin{align} \label{nut}
\nu_{t_0}^{\omega_1}=\mathcal{L}_2\Big(\int_0^{t_0} \sqrt{2\mathcal E b_\epsilon} (Z_{t_0}Z_s^{-1})(\omega_1)dB_s^{\omega_1}(\omega_2)\Big).
\end{align}
Here again we do not write the dependence in $\epsilon$ of $\mu_{t_0}^{\omega_1}$ and $\nu_{t_0}^{\omega_1}$.

- For $(\omega_1,\omega_2)\in\Omega$, we can now set $N(\omega_1,\omega_2)=N_{|\theta|\geq\epsilon}(\omega_1)+N_{|\theta|<\epsilon}^{\omega_1}(\omega_2)$ and $(B_t(\omega_1,\omega_2))_{t\geq0} =(B_t^{\omega_1}(\omega_2))_{t\geq0}$. Clearly, as random objects on $(\Omega,\mathcal{F},\mathcal{F}_t,\mathbb{P})$, the process $(B_t)_{t\geq0}$ is a $(\mathcal{F}_t)_{t\geq0}$-Brownian motion and $N$ is a $(\mathcal{F}_t)_{t\geq0}$-Poisson measure on $[0,\infty)\times[0,1]\times[-\pi,\pi]$ with intensity measure $dsd\alpha\beta(\theta)d\theta$. 

- Setting $\mathcal E:=\mathbb E[V_0^2]$, for $0<\epsilon<1$, we consider the processes $(V_t)_{t\geq0}$, $(V_t^\epsilon)_{t\geq0}$ defined on $(\Omega,\mathcal{F}, (\mathcal{F}_t)_{t\geq0}, \mathbb P)$ solutions to (\ref{eds}) and (\ref{Veps}) with $B$, $N$, $W$, $W^\epsilon$ defined previously, both starting from $V_0$. We also introduce the process $(\tilde V_t^\epsilon)_{t\geq0}$ solution of the following S.D.E.:
\begin{align} \label{Vtilde}
\tilde V_t^\epsilon=&V_0+\int_0^t\int_0^1\int_{|\theta|\geq\epsilon} \Big[(\cos\theta-1)\tilde V_{s-}^\epsilon-\sin\theta W_{s-}(\alpha)\Big]N(dsd\alpha d\theta)\\
\nonumber
&-\int_0^t\int_0^1\int_{|\theta|<\epsilon} \sin\theta W_{s-}(\alpha)\tilde N(dsd\alpha d\theta) -b_\epsilon\int_0^t \tilde V_s^\epsilon ds.
\end{align}
By Proposition \ref{def} and Proposition \ref{loi}, $\mathcal L(V_t)$ and $\mathcal L(V_t^\epsilon)$ are nothing but $f_t$ and $f_t^\epsilon$ respectively. We set $\Delta_t^\epsilon=V_t-\tilde V_t^\epsilon$, $\tilde\Delta_t^\epsilon=\tilde V_t^\epsilon-V_t^\epsilon$ and $\delta_t^\epsilon(\alpha)=W_s(\alpha)-W_s^\epsilon(\alpha)$.
\vskip0.5cm

\underline{Step 2}: the aim is here to prove that 
\begin{align} \label{step2}
\mathbb{E}\Big((\tilde\Delta_{t_0}^\epsilon)^2\Big)\leq d_\epsilon e^{-c_\epsilon t_0}\int_0^{t_0} e^{c_\epsilon s}\mathbb{E}_\alpha(\delta_s^2)ds+C\epsilon^2,
\end{align}
where $C$ depends only on $\mathcal E$ and $\mathbb E[V_0^4]$ and where $c_\epsilon$ and $d_\epsilon$ are defined in (\ref{ceps}). Making the difference between (\ref{Vtilde}) and (\ref{Veps}), we get
\begin{align} \label{tildedeltat}
\tilde\Delta_t^\epsilon&=\int_0^t\int_0^1\int_{|\theta|\geq\epsilon} \Big[(\cos\theta-1)\tilde\Delta_{s-}^\epsilon-\sin\theta \delta_{s-}^\epsilon(\alpha)\Big] N(dsd\alpha d\theta)-b_\epsilon\int_0^t \tilde\Delta_s^\epsilon ds\\
\nonumber
&\ \ \ \ -\int_0^t\int_0^1\int_{|\theta|<\epsilon} \sin\theta W_{s-}(\alpha)\tilde N(dsd\alpha d\theta)-\sqrt{2\mathcal E b_\epsilon} B_t\\
\nonumber
&=H_t+\int_0^t \tilde\Delta_{s-}^\epsilon dX_s,
\end{align}
with $(X_t)_{t\geq0}$ defined in (\ref{X}) and with
\begin{align*}
H_t=&-\int_0^t\int_0^1\int_{|\theta|\geq\epsilon} \sin\theta \delta_{s-}^\epsilon(\alpha)N(ds d\alpha d\theta)\\
& -\int_0^t\int_0^1\int_{|\theta|<\epsilon} \sin\theta W_{s-}(\alpha)\tilde N(dsd\alpha d\theta)-\sqrt{2\mathcal E b_\epsilon} B_t.
\end{align*}
We do not write the dependence in $\epsilon$ for $H$.
According to Jacod \cite{JACOD}, $\tilde\Delta_t^\epsilon=(L_t+D_t)Z_t$, where $Z_t$ was defined in (\ref{doleans}) and where
\begin{align} \label{D}
D_t=-\int_0^t\int_0^1\int_{|\theta|<\epsilon} Z_{s-}^{-1}\sin\theta W_{s-}(\alpha)\tilde N(dsd\alpha d\theta)-\sqrt{2\mathcal E b_\epsilon}\int_0^t Z_{s}^{-1} dB_s
\end{align}
and
\begin{align}
L_t=-\int_0^t\int_0^1\int_{|\theta|\geq\epsilon} Z_{s-}^{-1}\frac{\sin\theta}{\cos\theta} \delta_{s-}^\epsilon(\alpha)N(dsd\alpha d\theta).
\end{align}
To verify this, it suffices to apply It\^o's formula and observe that the process $((L_t+D_t)Z_t)_{t\geq0}$ satisfies the same S.D.E. than $(\tilde\Delta_t^\epsilon)_{t\geq0}$, i.e $(L_t+D_t)Z_t=H_t+\int_0^t(L_{s-}+D_{s-})Z_{s-}dX_s$. This S.D.E. has Lipschitz coefficients and thus has a unique solution.
The processes $(D_t)_{t\geq0}$ and $(L_t)_{t\geq0}$ depend on $\epsilon$ but we do not write this dependence.

Hence
\begin{align} \label{no1}
\mathbb{E}[(\tilde\Delta_t^\epsilon)^2]=\mathbb{E}[L_t^2Z_t^2]+\mathbb{E}[D_t^2Z_t^2]+2\mathbb{E}[L_tD_tZ_t^2].
\end{align}

- First, 
\begin{align} \label{no2}
\mathbb{E}[L_tD_tZ_t^2]=\mathbb{E}_1[\mathbb{E}_2(L_tD_tZ_t^2)]=\mathbb{E}_1[L_tZ_t^2\mathbb{E}_2(D_t)]=0,
\end{align}
because for $\omega=(\omega_1,\omega_2)$, we have $(L_tD_tZ_t^2)(\omega)=(L_tZ_t^2)(\omega_1)D_t(\omega_1,\omega_2)$ and because for $\omega_1$ fixed, $\mathbb E_2[D_t(\omega_1,\omega_2)]=0$. Indeed, recall that $(L_t)_{t\geq0}$ and $(Z_t)_{t\geq0}$ depend only on $\omega_1$ and that for $\omega_1$ fixed, $N_{|\theta|<\epsilon}(\omega_1,\omega_2)$ is a Poisson measure while $\big(B_t(\omega_1,\omega_2)\big)_{t\geq0}$ is a Brownian motion on $(\Omega_2,\mathcal{F}^2,\mathcal{F}_t^2,\mathbb{P}_2)$, so that $\big(D_t(\omega_1,\omega_2)\big)_{t\geq0}$ is a centered martingale (for $\omega_1$ fixed).

- By It\^o's formula, we have
\begin{align*}
Z_t^2L_t^2&=-2b_\epsilon\int_0^tZ_s^2L_s^2ds \\
&\ \ \  +\int_0^t\int_0^1\int_{|\theta|\geq\epsilon}\Big[\Big(Z_{s-}+(\cos\theta-1)Z_{s-}\Big)^2\Big(L_{s-}-\frac{\sin\theta}{\cos\theta}Z_{s-}^{-1}\delta_{s-}^\epsilon(\alpha)\Big)^2\\
&\ \ \ \ \ \ \ \ \ \ \ \ \ \ \ \ \ \ \ \ \ \ \ \ -Z_{s-}^2L_{s-}^2\Big]N(dsd\alpha d\theta)\\
&=-2b_\epsilon\int_0^tZ_s^2L_s^2ds \\
&\ \ \ +\int_0^t\int_0^1\int_{|\theta|\geq\epsilon}\Big[(\cos^2\theta-1)Z_{s-}^2L_{s-}^2+\sin^2\theta(\delta_{s-}^\epsilon(\alpha))^2\\
&\ \ \ \ \ \ \ \ \ \ \ \ \ \ \ \ \ \ \ \ \ \ \ \ \ -2\cos\theta\sin\theta Z_{s-}L_{s-}\delta_{s-}^\epsilon(\alpha)\Big]N(dsd\alpha d\theta).
\end{align*}
Taking expectations and recalling (\ref{ceps}), we get (use that $\cos\theta\sin\theta\beta(\theta)d\theta$ is odd)
$$\mathbb{E}(Z_t^2L_t^2)=-c_\epsilon\int_0^t\mathbb{E}(Z_s^2L_s^2)ds+d_\epsilon\int_0^t \mathbb{E}_\alpha[(\delta_s^\epsilon)^2]ds.$$
Solving this differential equation, we find 
\begin{align} \label{no3}
\mathbb{E}(Z_t^2L_t^2)=d_\epsilon e^{-c_\epsilon t}\int_0^t e^{c_\epsilon s}\mathbb{E}_\alpha[(\delta_s^\epsilon)^2]ds.
\end{align}

- It remains to compute $\mathbb{E}(Z_{t_0}^2D_{t_0}^2)$. Recalling (\ref{D}), we directly obtain
\begin{align*}
\mathbb{E}(Z_{t_0}^2D_{t_0}^2)&=\mathbb{E}\Big[\Big(Z_{t_0}\int_0^{t_0}\int_0^1\int_{|\theta|<\epsilon} Z_{s-}^{-1}\sin(-\theta)W_{s-}(\alpha)\tilde N(dsd\alpha d\theta)\\
&\ \ \ \ \ \ \ \ -Z_{t_0}\int_0^{t_0} Z_s^{-1}\sqrt{2\mathcal E b_\epsilon} dB_s\Big)^2\Big]\\
&=\mathbb{E}_1\Bigg[\mathbb{E}_2\Big[\Big(\int_0^{t_0}\int_0^1\int_{-\pi}^\pi (Z_{t_0}Z_{s-}^{-1})(\omega_1)\sin(-\theta)W_{s-}(\alpha)\tilde N_{|\theta|<\epsilon}^{\omega_1}(\omega_2,dsd\alpha d\theta)\\
&\ \ \ \ \ \ \ \ -\int_0^{t_0} (Z_{t_0}Z_s^{-1})(\omega_1)\sqrt{2\mathcal E b_\epsilon} dB_s^{\omega_1}(\omega_2)\Big)^2\Big]\Bigg].
\end{align*}
We thus obtain $\mathbb{E}(Z_{t_0}^2D_{t_0}^2)=\mathbb{E}_1(W_2^2(\mu_{t_0},\nu_{t_0}))$, recall (\ref{wasser}). We consider
\begin{align} \label{etat}
\eta_{t_0}^{\omega_1}:=\mathcal{L}_2\Big( \sigma_\epsilon\int_0^{t_0} (Z_{t_0}Z_s^{-1})(\omega_1)dB_s^{\omega_1}(\omega_2)\Big),
\end{align}
with
$$ \sigma_\epsilon=\sqrt{\mathcal{E}\int_{|\theta|<\epsilon} \sin^2\theta \beta(\theta)d\theta}.$$
Using the triangular inequality, we have
$$W_2^2(\mu_{t_0}^{\omega_1},\nu_{t_0}^{\omega_1})\leq 2(W_2^2(\mu_{t_0}^{\omega_1},\eta_{t_0}^{\omega_1})+W_2^2(\eta_{t_0}^{\omega_1},\nu_{t_0}^{\omega_1})).$$
By Corollary \ref{coro} and since $\mathbb{E}_\alpha(W_s^2)=\mathcal E$ for all $s\geq 0$ by the energy conservation, we have (recall that $\omega_1$ is fixed)
$$W_2^2(\mu_{t_0}^{\omega_1},\eta_{t_0}^{\omega_1})\leq C_0\frac{\int_0^{t_0}\int_{|\theta|<\epsilon}(Z_{t_0}Z_s^{-1})^4(\omega_1)\sin^4\theta\mathbb{E}_\alpha(W_s^4)\beta(\theta)d\theta ds}{\int_0^{t_0}\int_{|\theta|<\epsilon}(Z_{t_0}Z_s^{-1})^2(\omega_1)\sin^2\theta\mathcal{E}\beta(\theta)d\theta ds}.$$
But $\mathbb{E}_\alpha(W_s^4)\leq\mathbb{E}(V_0^4)+3\mathcal{E}^2$ by Lemma \ref{moment4}. Furthermore, recalling (\ref{doleans}), we have $|Z_{t_0}Z_s^{-1}|=|e^{-b_\epsilon(t_0-s)}\prod_{s\leq T_i\leq t}\cos\theta_i|\leq1$. Since finally $\sin^2\theta\leq\theta\leq\epsilon^2$ on $[-\epsilon,\epsilon]$, we easily deduce that for all $\omega_1$ fixed,
$$W_2^2(\mu_{t_0}^{\omega_1},\eta_{t_0}^{\omega_1})\leq\frac{C_0(\mathbb{E}(V_0^4)+3\mathcal{E}^2)}{\mathcal{E}}\epsilon^2.$$
Finally, it obviously holds, recall (\ref{nut}) and (\ref{etat}), that for all $\omega_1$ fixed,
\begin{align*}
W_2^2(\eta_{t_0}^{\omega_1},\nu_{t_0}^{\omega_1})&\leq  \int_0^{t_0}\Big(\sqrt{2\mathcal E b_\epsilon}-\sqrt{\mathcal E \int_{|\theta|<\epsilon}\sin^2\theta \beta(\theta)d\theta}\Big)^2(Z_s^{-1}Z_{t_0})^2(\omega_1)ds\\
&\leq \mathcal E\Big|2b_\epsilon-\int_{|\theta|<\epsilon}\sin^2\theta \beta(\theta)d\theta\Big|\int_0^{t_0}(Z_s^{-1}Z_{t_0})^2(\omega_1)ds.
\end{align*}
We used that $(\sqrt x-\sqrt y)^2\leq |x-y|$. Recalling that $|Z_s^{-1}Z_{t_0}|\leq e^{-b_\epsilon(t_0-s)}$, we easily get $\int_0^{t_0}(Z_s^{-1}Z_{t_0})^2(\omega_1)ds\leq\frac{1}{2b_\epsilon}$. Furthermore, $\Big|2b_\epsilon-\int_{|\theta|<\epsilon}\sin^2\theta \beta(\theta)d\theta\Big|=\Big|\int_{|\theta|<\epsilon}\big(2(1-\cos\theta)-\sin^2\theta\big) \beta(\theta)d\theta\Big|\leq\int_{|\theta|<\epsilon}\theta^4 \beta(\theta)d\theta$. Finally, it is easily checked that, since $\epsilon\in(0,1)$, $b_\epsilon=\int_{|\theta|<\epsilon}(1-\cos\theta) \beta(\theta)d\theta\geq\frac{1}{4}\int_{|\theta|<\epsilon}\theta^2 \beta(\theta)d\theta$. Hence it holds that for all $\omega_1$ fixed,
\begin{align*}
W_2^2(\eta_{t_0}^{\omega_1},\nu_{t_0}^{\omega_1})\leq  2\mathcal E\frac{\int_{|\theta|<\epsilon}\theta^4 \beta(\theta)d\theta}{\int_{|\theta|<\epsilon}\theta^2 \beta(\theta)d\theta}\leq2\mathcal E\epsilon^2.
\end{align*}
We conclude that $W_2^2(\mu_{t_0}^{\omega_1},\nu_{t_0}^{\omega_1})\leq C\epsilon^2$ (where $C$ depends on $\mathcal E$ and $\mathbb E[V_0^4]$), whence 
\begin{align} \label{no4}
\mathbb E[Z_{t_0}^2D_{t_0}^2]\leq C\epsilon^2.
\end{align}
Gathering (\ref{no1}), (\ref{no2}), (\ref{no3}) and (\ref{no4}), we deduce (\ref{step2}).
\vskip0.5cm

\underline{Step 3}: in this step, we check that 
\begin{align} \label{step3}
\mathbb{E}[(\Delta_t^\epsilon)^2]\leq\frac{\mathcal E}{4}\epsilon^2
\end{align}
for all $t\geq0$. We first observe that (\ref{eds}) can be rewritten as
\begin{align*}
V_t=&V_0+\int_0^t\int_0^1\int_{|\theta|\geq\epsilon} \Big[(\cos\theta-1) V_{s-}-\sin\theta W_{s-}(\alpha)\Big]N(dsd\alpha d\theta)\\ &+\int_0^t\int_0^1\int_{|\theta|<\epsilon} \Big[(\cos\theta-1)V_{s-}-\sin\theta W_{s-}(\alpha)\Big]\tilde N(dsd\alpha d\theta) -b_\epsilon\int_0^t V_s ds.
\end{align*}
Hence, making the difference with (\ref{Vtilde}), we find
\begin{align} \label{deltat}
\Delta_t^\epsilon=&\int_0^t\int_0^1\int_{|\theta|\geq\epsilon} (\cos\theta-1) \Delta_{s-}^\epsilon N(dsd\alpha d\theta)\\
\nonumber
&+\int_0^t\int_0^1\int_{|\theta|<\epsilon} (\cos\theta-1)V_{s-}\tilde N(dsd\alpha d\theta) -b_\epsilon\int_0^t \Delta_s^\epsilon ds.
\end{align}
Applying It\^o's formula, we get
\begin{align*}
(\Delta_t^\epsilon)^2&=-2b_\epsilon\int_0^t(\Delta_s^\epsilon)^2ds+\int_0^t\int_0^1\int_{|\theta|\geq\epsilon} (\cos^2\theta-1) (\Delta_{s-}^\epsilon)^2N(dsd\alpha d\theta)\\ 
&\quad+\int_0^t\int_0^1\int_{|\theta|<\epsilon} \Big[\Big(\Delta_{s-}^\epsilon+(\cos\theta-1)V_{s-}\Big)^2-(\Delta_{s-}^\epsilon)^2\Big]\tilde N(dsd\alpha d\theta)\\
&\quad+\int_0^t\int_0^1\int_{|\theta|<\epsilon} \Big[\Big(\Delta_{s}^\epsilon+(\cos\theta-1)V_{s}\Big)^2-(\Delta_{s}^\epsilon)^2\\
&\quad\quad\quad\quad\quad\quad\quad\quad-2\Delta_s^\epsilon(\cos\theta-1)V_s\Big]\beta(\theta)dsd\alpha d\theta\\
&=-\Big(2b_\epsilon\int_0^t(\Delta_s^\epsilon)^2ds+\int_0^t\int_0^1\int_{|\theta|\geq\epsilon} \sin^2\theta (\Delta_{s-}^\epsilon)^2N(dsd\alpha d\theta)\Big)+M_t\\
&\quad+\int_0^t\int_{|\theta|<\epsilon}V_s^2(1-\cos\theta)^2\beta(\theta)d\theta ds,
\end{align*}
where $(M_t)_{t\geq0}$ is a centered matingale. Taking expectations, this yields, recalling (\ref{ceps}) and that $\mathbb E[V_s^2]=\mathcal E$ for all $s\geq0$ by the energy conservation,
\begin{align*}
\mathbb{E}[(\Delta_t^\epsilon)^2]&=-c_\epsilon\int_0^t\mathbb{E}[(\Delta_s^\epsilon)^2]ds+\gamma_\epsilon\mathcal{E}t.
\end{align*}
Thus
$$\mathbb{E}[(\Delta_t^\epsilon)^2]=\frac{\gamma_\epsilon\mathcal{E}}{c_\epsilon}(1-e^{-c_\epsilon t})\leq \frac{\gamma_\epsilon\mathcal{E}}{c_\epsilon}.$$
But $\gamma_\epsilon\leq b_\epsilon\frac{\epsilon^2}{2}$ (because for $|\theta|<\epsilon$, $(1-\cos\theta)\leq\theta^2/2<\epsilon^2/2)$  and $c_\epsilon\geq2b_\epsilon$. We deduce that $\gamma_\epsilon/c_\epsilon\leq \epsilon^2/4$ and finally get (\ref{step3}).
\vskip0.5cm

\underline{Step 4}: we now check that $\mathbb{E}[\Delta_t^\epsilon\tilde\Delta_t^\epsilon]=0$ for all $t\geq0$.
Applying It\^o's formula, using (\ref{deltat}) and (\ref{tildedeltat}), we have
\begin{align*}
\Delta_t^\epsilon\tilde\Delta_t^\epsilon=&-\sqrt{2\mathcal E b_\epsilon}\int_0^t \Delta_s^\epsilon dB_s -2b_\epsilon\int_0^t \Delta_s^\epsilon\tilde\Delta_s^\epsilon ds\\
&+\int_0^t\int_0^1\int_{|\theta|\geq\epsilon} \Big[(\cos\theta\Delta_{s-}^\epsilon)(\cos\theta\tilde\Delta_{s-}^\epsilon-\sin\theta\delta_{s-}^\epsilon(\alpha))\\
&\ \ \ \ \ \ \ \ \ \ \ \ \ \ \ \ \ \ \ \ \ -\Delta_{s-}^\epsilon\tilde\Delta_{s-}^\epsilon\Big]N(ds d\alpha d\theta)\\
&+\int_0^t\int_0^1\int_{|\theta|<\epsilon} \Big[\Big(\Delta_{s-}^\epsilon+(\cos\theta-1)V_{s-}\Big)\Big(\tilde\Delta_{s-}^\epsilon-\sin\theta W_{s-}(\alpha)\Big)\\
&\ \ \ \ \ \ \ \ \ \ \ \ \ \ \ \ \ \ \ \ \ -\Delta_{s-}^\epsilon\tilde\Delta_{s-}^\epsilon\Big]\tilde N(ds d\alpha d\theta)\\
&+\int_0^t\int_0^1\int_{|\theta|<\epsilon} \Big[\Big(\Delta_{s}^\epsilon+(\cos\theta-1)V_{s}\Big)\Big(\tilde\Delta_{s}^\epsilon-\sin\theta W_{s}(\alpha)\Big)\\
&\ \ \ \ \ \ \ \ \ \ \ \ \ \ \ \ \ \ \ \ \ -\Delta_{s}^\epsilon\tilde\Delta_{s}^\epsilon-\tilde\Delta_s^\epsilon(\cos\theta-1)V_s+\Delta_s^\epsilon\sin\theta W_s(\alpha)\Big]\beta(\theta)ds d\alpha d\theta.
\end{align*}
Taking expectation and using that $\beta$ is even, we get
\begin{align*}
\mathbb{E}[\Delta_t^\epsilon\tilde\Delta_t^\epsilon]=-2b_\epsilon\int_0^t \mathbb{E}[\Delta_s^\epsilon\tilde\Delta_s^\epsilon]ds+\int_0^t\int_{|\theta|\geq\epsilon}(\cos^2\theta-1)\mathbb{E}[\Delta_s^\epsilon\tilde\Delta_s^\epsilon]\beta(\theta)d\theta ds.
\end{align*}
So the function $t\mapsto \mathbb{E}[\Delta_t^\epsilon\tilde\Delta_t^\epsilon]$ solves the O.D.E. $y'=-c_\epsilon y$, see (\ref{ceps}). Since $y(0)=0$, we easily conclude.
\vskip0.5cm

\underline{Step 5}: conclusion. Using Steps 2, 3 and 4, we find that
\begin{align*}
\mathbb{E}[(V_{t_0}-V_{t_0}^\epsilon)^2]&=\mathbb{E}[(\Delta_{t_0}^\epsilon)^2]+\mathbb{E}[(\tilde\Delta_{t_0}^\epsilon)^2]+2\mathbb E[\Delta_{t_0}^\epsilon\tilde\Delta_{t_0}^\epsilon]\\
&\leq d_\epsilon e^{-c_\epsilon t_0}\int_0^{t_0}e^{c_\epsilon s}\mathbb{E}_\alpha[(\delta_s^\epsilon)^2]ds+K\epsilon^2,
\end{align*}
where $K$ depends only on $\mathcal E$ and $\mathbb E[V_0^4]$.
We set $u(t)=W_2^2(f_t,f_t^\epsilon)=\mathbb{E}_\alpha[(\delta_s^\epsilon)^2]$ by (\ref{delta}). Since $\mathcal L(V_{t_0})=f_{t_0}$ and $\mathcal L(V_{t_0}^\epsilon)=f_{t_0}^\epsilon$, we have $u(t_0)\leq\mathbb{E}[(V_{t_0}-V_{t_0}^\epsilon)^2]$. Since $t_0\geq0$ is arbitrary, we get, for all $t\geq0$, 
\begin{align*}
u(t)\leq d_\epsilon e^{-c_\epsilon t}\int_0^te^{c_\epsilon s}u(s)ds+K\epsilon^2=:v(t).
\end{align*}
Consequently,
\begin{align*}
v'(t)&=-c_\epsilon\Big(v(t)-K\epsilon^2\Big)+d_\epsilon u(t)\\
&\leq -c_\epsilon\Big(v(t)-K\epsilon^2\Big)+d_\epsilon v(t)\\
&\leq (d_\epsilon-c_\epsilon)v(t)+c_\epsilon K\epsilon^2.
\end{align*}
We first observe that $d_\epsilon\leq c_\epsilon$, so that $v(t)\leq v(0)+Kc_\epsilon\epsilon^2t\leq K\epsilon^2(1+c_\epsilon)(1+t)\leq C\epsilon^2(1+t)$, because $c_\epsilon\leq \int_{-\pi}^\pi\theta^2\beta(\theta)d\theta$, see (\ref{ceps}).
\vskip0.5cm

We can also obtain a uniform in time bound. Recall that $v'(t)\leq (d_\epsilon-c_\epsilon)v(t)+c_\epsilon K\epsilon^2=-2b_\epsilon v(t)+c_\epsilon K\epsilon^2$. We observe in fact that $v'(t)\leq0$ as soon as $v(t)\geq\frac{c_\epsilon K\epsilon^2}{2b_\epsilon}$. Since $v(0)=K\epsilon^2\leq\frac{c_\epsilon K\epsilon^2}{2b_\epsilon}$, we classically deduce that $v(t)\leq\frac{c_\epsilon K\epsilon^2}{2b_\epsilon}\leq C\frac{\epsilon^2}{b_\epsilon}$ for all $t\geq0$.

\vskip0.5cm
So we have $W_2^2(f_t,f_t^\epsilon)=u(t)\leq v(t)\leq C\min\Big(\epsilon^2(1+t),\frac{\epsilon^2}{b_\epsilon}\Big)$ for all $t\geq0$. To complete the proof of Theorem \ref{petits sauts}, it suffices to observe that $4b_\epsilon\geq\int_{|\theta|<\epsilon}\theta^2\beta(\theta)d\theta$ for any $\epsilon\in(0,1)$. \hfill$\square$

\section{Convergence of the particle system}

In this section, we prove the results about the approximation of the solution of the Kac equation by a system of particles. Let thus $f_0\in\mathcal P_4(\mathbb R)$ and let $\beta$ be a cross section satisfying (\ref{ncutoff}). We fix $\epsilon\in(0,1)$, and we consider the solutions $(f_t)_{t\geq0}$ and $(f_t^\epsilon)_{t\geq0}$ to (\ref{Kac}) and (\ref{mixte}) respectively, both starting from $f_0$.
\vskip0.5cm

In the first part, we will rewrite the system of particles (\ref{system}) in a suitable way and in the second part, we will introduce a system of i.i.d. particles with law $(f_t^\epsilon)_{t\geq0}$. Using these systems of particles, we will be able to prove Theorem \ref{systeme} and its corollary. We will end this section with the proof of Proposition \ref{W1} and with an extension about the Wasserstein distance $W_\gamma$ for $\gamma\in(1,2)$. 
\vskip0.5cm

We recall a usefull result (see e.g. Villani \cite[Remark 2.19 (iii)]{VIL2}).
\begin{prop} \label{Villani}
If $\mu$ and $\nu$ are two probability measures on $\mathbb R$, for $\gamma\geq1$, we have $W_\gamma^\gamma(\mu,\nu)=\int_0^1\big(F_\mu^{-1}(\alpha)-F_\nu^{-1}(\alpha)\big)^\gamma d\alpha$ where $F_\mu(x)=\mu\big((-\infty,x]\big)$ and $F_\nu(x)=\nu\big((-\infty,x]\big)$.
\end{prop}

\subsection{Another way to write system (\ref{system})}

We fix an integer $n$ and we consider:
\begin{itemize}
	\item a family of i.i.d. random variables $(V_0^i)_{i\in\{1,...,n\}}$ with law $f_0$,
	\item a family of i.i.d. Poisson measures $(N^{i})_{i\in\{1,...,n\}}$ on $[0,\infty)\times[0,1]\times[-\pi,\pi]$ with intensity measure $dtd\alpha\beta(\theta)d\theta$, 
	\item a family of i.i.d. Brownian motions $(B_t^i)_{t\geq0, i\in\{1,...,n\}}$.
\end{itemize}
For $\epsilon\in(0,1)$, we consider $(V_t^{i,n, \epsilon})_{t\geq0,\ i\in\{1,...,n\}}$ solution of the following system of S.D.E.s: for $i=1,...,n$, for all $t\geq0$,
\begin{align*}
V_t^{i,n, \epsilon}=&V_0^i+\int_0^t\int_0^1\int_{|\theta|>\epsilon} \Big[(\cos\theta-1)V_{s-}^{i,n, \epsilon}-\sin\theta\Big(F_{s-}^{n,\epsilon}\Big)^{-1}(\alpha)\Big]N^{i}(d\theta d\alpha ds)\\
& -b_\epsilon\int_0^t V_s^{i,n,\epsilon}ds + \sqrt{2\mathcal E b_\epsilon} B_t^i,
\end{align*}
where $F_t^{n,\epsilon}=\frac{1}{n}\sum_1^n \one_{V_t^{i,n, \epsilon}\leq x}$, $\mathcal E=\int_{\mathbb R} v^2f_0(dv)$ and $b_\epsilon=\int_{|\theta|<\epsilon}(1-\cos\theta)\beta(\theta)d\theta$. 
\vskip0.5cm

This particle system is identical (in law) to the one introduced in (\ref{system}). Indeed, it suffices to note that given $(V_{s-}^{i,n, \epsilon})_{ i\in\{1,...,n\}}$, the law of $(F_{s-}^{n,\epsilon})^{-1}(\alpha)$ (with $\alpha$ uniformly distributed on $[0,1]$) is the same as that of $V_{s-}^{j,n, \epsilon}$ (with $j$ uniformly distributed in $\{1,...,n\}$).  

\subsection{A system of i.i.d. particles}

For $i\in\{1,...,n\}$ and $\epsilon>0$, we consider the process $(\bar V_t^{i,\epsilon})_{t\geq0}$ solution of the following S.D.E. (with the same random objects $V_0^i$, $N^i$ and $B^i$ as previously),
\begin{align*}
\bar V_t^{i,\epsilon}=&V_0^i+\int_0^t\int_0^1\int_{|\theta|>\epsilon} \Big[(\cos\theta-1)\bar V_{s-}^{i,\epsilon}-\sin\theta\Big(F_{s-}^\epsilon\Big)^{-1}(\alpha)\Big]N^{i}(dsd\alpha d\theta)\\
& -b_\epsilon\int_0^t \bar V_s^{i,\epsilon}ds + \sqrt{2\mathcal E b_\epsilon} B_t^i,
\end{align*}
where $F_t^\epsilon(x)=\int_{-\infty}^x f_t^\epsilon(dv)$.

For each $s\geq0$, it holds that $\mathcal L_\alpha\big((F_s^\epsilon)^{-1}\big))=f_s^\epsilon$. Hence we can apply Proposition \ref{loi} with $W_s^\epsilon=(F_s^\epsilon)^{-1}$ and deduce that for each $i\in\{1,...,n\}$, each $t\geq0$, $\mathcal L(\bar V_t^{i,\epsilon})=f_t^\epsilon$. Furthermore, the processes $(\bar V_t^i)_{t\geq0}$ are obviously i.i.d. (for $i=1,...,n$).

\subsection{Proof of Theorem \ref{systeme}}

We start with the following result.

\begin{prop} \label{particules} 
We set $\mu_t^{n,\epsilon}=\frac{1}{n}\sum_1^n \delta_{V_t^{i,n, \epsilon}}$ and $\bar\mu_t^{n,\epsilon}=\frac{1}{n}\sum_1^n \delta_{\bar V_t^{i,\epsilon}}$. Then for any $T>0$,
$$\sup_{[0,T]}\mathbb E\Big[W_2^2(f_t^\epsilon,\mu_t^{n,\epsilon})\Big]\leq C(1+T)^2 \sup_{[0,T]}\mathbb E\Big[W_2^2(f_t^\epsilon,\bar\mu_t^{n,\epsilon})\Big],$$
where $C$ depends only on $\int_0^\pi \theta^2\beta(\theta)d\theta$.
\end{prop}

\textbf{Proof.}
To lighten notation, we set $V_t^{i}= V_t^{i,n,\epsilon}$ and $\bar V_t^{i}=\bar V_t^{i,\epsilon}$ for the whole proof. By the triangular inequality, we have $W_2(f_t^\epsilon,\mu_t^{n,\epsilon})\leq W_2(f_t^\epsilon,\bar\mu_t^{n,\epsilon})+W_2(\bar\mu_t^{n,\epsilon},\mu_t^{n,\epsilon})$. Hence, by squaring and taking expectations
\begin{align} \label{particule}
\mathbb E\Big[W_2^2(f_t^\epsilon,\mu_t^{n,\epsilon})\Big]\leq & \mathbb E\Big[W_2^2(f_t^\epsilon,\bar\mu_t^{n,\epsilon})\Big]+\mathbb E\Big[W_2^2(\bar\mu_t^{n,\epsilon},\mu_t^{n,\epsilon})\Big]\\ 
&+2\mathbb E\Big[W_2(f_t^\epsilon,\bar\mu_t^{n,\epsilon})W_2(\bar\mu_t^{n,\epsilon},\mu_t^{n,\epsilon})\Big]. \nonumber
\end{align}
Using the fact that $W_2^2\Big(\frac{1}{n}\sum_1^n\delta_{x_i},\frac{1}{n}\sum_1^n\delta_{y_i}\Big)\leq \frac{1}{n}\sum_1^n|x_i-y_i|^2$, we have
\begin{align*}
\mathbb E\Big[W_2^2(\bar\mu_t^{n,\epsilon},\mu_t^{n,\epsilon})\Big]\leq\mathbb E\Big(\frac{1}{n}\sum_{i=1}^{n}|\bar V_t^i-V_t^i|^2\Big)=\mathbb E\Big(|\bar V_t^1-V_t^1|^2\Big).
\end{align*}
We set $\Delta_t=\bar V_t^1-V_t^1$. It holds that
\begin{align*}
\Delta_t=\int_0^t\int_0^1\int_{|\theta|>\epsilon} \Big[(\cos\theta-1)\Delta_{s-}-\sin\theta\delta_{s-}(\alpha)\Big]N^{1}(dsd\alpha d\theta)-b_\epsilon \int_0^t \Delta_sds,
\end{align*}
where $\delta_t(\alpha)=\big( F_t^{n,\epsilon}\big)^{-1}(\alpha)-\big(F_t^\epsilon\big)^{-1}(\alpha).$
Applying It\^o's formula, we get
\begin{align*}
\Delta_t^2=&\int_0^t\int_0^1\int_{|\theta|>\epsilon} \Big[\big(\Delta_s+(\cos\theta-1)\Delta_{s}-\sin\theta\delta_{s}(\alpha)\big)^2-\Delta_s^2\Big]N^{1}(dsd\alpha d\theta)\\
&-2b_\epsilon \int_0^t \Delta_s^2ds.
\end{align*}
Taking expectations and using Proposition \ref{Villani}, we get, with $c_\epsilon$ and $d_\epsilon$ defined in (\ref{ceps}),
\begin{align*}
v(t):=\mathbb E(\Delta_t^2)&=-c_\epsilon\int_0^t \mathbb E(\Delta_s^2)ds+d_\epsilon\int_0^t\mathbb E\Big(\int_0^1 \delta_s^2(\alpha)d\alpha\Big)ds\\
&=-c_\epsilon\int_0^t v(s)ds+d_\epsilon\int_0^t\mathbb E[W_2^2(f_s^\epsilon,\mu_s^{n,\epsilon})]ds\\
&=d_\epsilon e^{-c_\epsilon t}\int_0^t e^{c_\epsilon s}\mathbb E[W_2^2(f_s^\epsilon,\mu_s^{n,\epsilon})]ds,
\end{align*}
the last equality being obtained by solving the differential equation satisfied by $v$. 
If we set $u(t):=\mathbb E\Big[W_2^2(f_t^\epsilon,\mu_t^{n,\epsilon})\Big]$, $s_n:=\sup_{[0,T]}\mathbb E\Big[W_2^2(f_s^\epsilon,\bar\mu_s^{n,\epsilon})\Big]$ and if we return to (\ref{particule}), we thus find, for all $t\in[0,T]$,
\begin{align*}
u(t)&\leq s_n+v(t)+2\mathbb E\Big[W_2(f_t^\epsilon,\bar\mu_t^n)W_2(\bar\mu_t^{n,\epsilon},\mu_t^{n,\epsilon})\Big]\\
&\leq s_n+v(t)+2\sqrt{s_n}\sqrt{v(t)},
\end{align*}
by the Cauchy-Schwarz inequality. We thus have
\begin{align*}
v(t)&=d_\epsilon e^{-c_\epsilon t}\int_0^t e^{c_\epsilon s}u(s)ds\\
&\leq d_\epsilon e^{-c_\epsilon t}\int_0^t e^{c_\epsilon s}\Big(s_n+v(s)+2\sqrt{s_n}\sqrt{v(s)}\Big)ds\\
&\leq s_n+d_\epsilon e^{-c_\epsilon t}\int_0^t e^{c_\epsilon s}\Big(v(s)+2\sqrt{s_n}\sqrt{v(s)}\Big)ds=:w(t).
\end{align*}
We used that $d_\epsilon e^{-c_\epsilon t}\int_0^t e^{c_\epsilon s}ds=\frac{d_\epsilon}{c_\epsilon}(1-e^{-c_\epsilon t})\leq1$, since $d_\epsilon\leq c_\epsilon$, recall (\ref{ceps}).
Differentiating $w$, we get
\begin{align*}
w'(t)&=-c_\epsilon\big(w(t)-s_n\big)+d_\epsilon\big(v(t)+2\sqrt{s_n}\sqrt{v(t)}\big)\\
&\leq c_\epsilon s_n+w(t)(-c_\epsilon+d_\epsilon)+2d_\epsilon\sqrt{s_n}\sqrt{w(t)}\\
&\leq c_\epsilon s_n+2d_\epsilon\sqrt{s_n}\sqrt{w(t)}\\
&\leq a s_n+2a\sqrt{s_n}\sqrt{w(t)},
\end{align*}
where $a=\int_{-\pi}^\pi\theta^2\beta(\theta)d\theta$.

Putting $x(t)=w(t)/s_n$, we deduce that $x(0)=1$ and $x'(t)\leq a(1+2\sqrt{x(t)})\leq2a(1+\sqrt{x(t)})\leq4a\sqrt{1+x(t)}$, whence $\sqrt{1+x(t)}-\sqrt{1+x(0)}\leq2at$, which gives $x(t)\leq C(1+t)^2$, and so $w(t)\leq C(1+t)^2s_n$.
\vskip0.5cm

To summarize, we have $v(t)\leq w(t)\leq C s_n(1+t)^2$ and, for all $t\in[0,T]$,
\begin{align*}
\mathbb E\Big[W_2^2(f_t^\epsilon,\mu_t^{n,\epsilon})\Big]=u(t)&\leq s_n+v(t)+2\sqrt{s_n}\sqrt{v(t)}\\
&\leq s_n+C s_n(1+t)^2+2\sqrt{s_n}\sqrt C \sqrt{s_n}(1+t)\\
&\leq Cs_n(1+t)^2\\
&=C\sup_{[0,T]}\mathbb E\Big[W_2^2(f_s^\epsilon,\bar\mu_s^{n,\epsilon})\Big](1+t)^2.
\end{align*}
This concludes the proof. \hfill$\square$
\vskip1cm

Theorem \ref{systeme} follows almost immediately.
\vskip0.5cm

\textbf{Proof of Theorem \ref{systeme}.} 
For  each $t\geq0$, we consider an i.i.d. sequence $(\hat V_t^i)_{i\in\{1,...,n\}}$ with law $f_t$ such that for each $i$, $\mathbb E[(\hat V_t^i-\bar V_t^{i,\epsilon})^2]=W_2^2(f_t,f_t^\epsilon)$. Then we set $\mu_t^{n}=\frac{1}{n}\sum_1^n \delta_{\hat V_t^i}$.
Using the triangular inequality, Theorem \ref{petits sauts} and Proposition \ref{particules}, we have
\begin{align}
\nonumber
\sup_{[0,T]}\mathbb E\Big[W_2^2(f_t,\mu_t^{n,\epsilon})\Big]&\leq 2\sup_{[0,T]}W_2^2(f_t,f_t^\epsilon)+2\sup_{[0,T]}\mathbb E\Big[W_2^2(f_t^\epsilon,\mu_t^{n,\epsilon})\Big]\\
\label{theo2.7}
&\leq C(1+T)\epsilon^2+C(1+T)^2 \sup_{[0,T]}\mathbb E\Big[W_2^2(f_t^\epsilon,\bar\mu_t^{n,\epsilon})\Big].
\end{align}
We use again the triangular inequality to obtain
\begin{align*}
\mathbb E\Big[W_2^2(f_t^\epsilon,\bar\mu_t^{n,\epsilon})\Big]&\leq 4\Big(W_2^2(f_t^\epsilon,f_t)+\mathbb E\Big[W_2^2(f_t,\mu_t^n)\Big]+\mathbb E\Big[W_2^2(\mu_t^n,\bar\mu_t^{n,\epsilon})\Big]\Big).
\end{align*}
But $\mathbb E\Big[W_2^2(\mu_t^n,\bar\mu_t^{n,\epsilon})\Big]\leq\frac{1}{n}\sum_{i=1}^n\mathbb E\Big(|\hat V_t^i- \bar V_t^{i,\epsilon}|^2\Big)= W_2^2(f_t,f_t^\epsilon)$. So using Theorem \ref{petits sauts}, we get
\begin{align} \label{theo2.72}
\sup_{[0,T]}\mathbb E\Big[W_2^2(f_t^\epsilon,\bar\mu_t^{n,\epsilon})\Big]&\leq C\Big((1+T)\epsilon^2+\sup_{[0,T]}\mathbb E\Big[W_2^2(f_t,\mu_t^n)\Big]\Big).
\end{align}
Inserting (\ref{theo2.72}) in (\ref{theo2.7}), we obtain
\begin{align*}
\sup_{[0,T]}\mathbb E\Big[W_2^2(f_t,\mu_t^{n,\epsilon})\Big]\leq C(1+T)^3\Big(\epsilon^2+\sup_{[0,T]}\mathbb E\Big[W_2^2(f_t,\mu_t^n)\Big]\Big),
\end{align*}
which concludes the proof. \hfill$\square$
\vskip0.5cm

Finally, we give the proof of Corollary \ref{corosysteme}.
\vskip0.5cm

\textbf{Proof of Corollary \ref{corosysteme}}
It suffices to apply Theorem \ref{systeme}, Lemma \ref{empirique} with $\gamma=2$ and $q=p-\gamma$, and Lemma \ref{moment4}. \hfill$\square$

\subsection{Other Wasserstein distances}

The first part of the following result is Proposition \ref{W1} and in the second part, we give some estimates about $\mathbb E[W_\gamma^\gamma(f_t,\mu_t^{n,\epsilon})]$ for $\gamma\in(1,2)$.

\begin{prop} \label{rate}
Adopt the same notation as in Theorem \ref{systeme}.

(i) If we assume that $f_0\in\mathcal P_4(\mathbb R)$ and if $\int_0^\pi \theta\beta(\theta)d\theta<\infty$, then 
\begin{align*}
\sup_{[0,T]}\mathbb E[W_1(f_t,\mu_t^{n,\epsilon})]\leq C_T\Big(\epsilon+\frac{1}{\sqrt n}\Big),
\end{align*}
where $C_T$ depends only on $T$, $f_0$ and $\beta$.

(ii) If $f_0\in\mathcal P_p(\mathbb R)$ for  some even $p\geq4$ and if $\int_0^\pi \theta^\gamma\beta(\theta)d\theta<\infty$ for some $\gamma\in(1,2)$, then
\begin{align*}
\sup_{[0,T]}\mathbb E[W_\gamma^\gamma(f_t,\mu_t^{n,\epsilon})]\leq C_T\Big(\epsilon^\gamma+\frac{1}{n^{\frac{p-\gamma}{2p-2}}}\Big),
\end{align*} 
where $C_T$ depends only on $T$, $f_0$, $\beta$, $p$ and $\gamma$.
\end{prop}

\textbf{Proof.}
Let $\gamma\in[1,2)$ be fixed. We assuùme that $\int_0^\pi \theta^\gamma\beta(\theta)d\theta<\infty$ and we set $V_t^{i}= V_t^{i,n,\epsilon}$ and $\bar V_t^{i}=\bar V_t^{i,\epsilon}$ for the whole proof to lighten notation.

\underline{Step 1}: we first prove that
\begin{align*}
I_{\gamma,\epsilon}(x):=\int_{|\theta|>\epsilon} \big[|\cos\theta-x\sin\theta|^\gamma-1\big]\beta(\theta)d\theta\leq C(1+|x|^\gamma).
\end{align*}

To this end, we set $J_{\gamma,\epsilon}(x)=\int_{|\theta|>\epsilon} \big[|1-x\theta|^\gamma-1\big]\beta(\theta)d\theta$. Using the inequality $|a^\gamma-b^\gamma|\leq C|a-b|(a^{\gamma-1}+b^{\gamma-1})$, we get
	\begin{align*}
	|I_{\gamma,\epsilon}(x)-J_{\gamma,\epsilon}(x)|&\leq C\int_{|\theta|>\epsilon}|(\cos\theta-1)-x(\sin\theta-\theta)|\\
	&\ \ \ \ \ \ \ \ \ \ \ \ \ \ \big(|\cos\theta-x\sin\theta|^{\gamma-1}+|1-x\theta|^{\gamma-1}\big)\beta(\theta)d\theta\\
	&\leq C\int_{|\theta|>\epsilon}\theta^2(1+|x|)(1+|x|^{\gamma-1})\beta(\theta)d\theta\\
	&\leq C(1+|x|)(1+|x|^{\gamma-1})\\
	&\leq C(1+|x|^\gamma).
	\end{align*}
	Using the fact that $\beta$ is even, we can write
	$$J_{\gamma,\epsilon}(x)=\int_{|\theta|>\epsilon}[|1-x\theta|^\gamma-1+\gamma x\theta]\beta(\theta)d\theta=J_{\gamma,\epsilon}^1(x)+J_{\gamma,\epsilon}^2(x)$$
	with
	$$J_{\gamma,\epsilon}^1(x)=\int_{|\theta|>\epsilon,|x\theta|<1/2}[|1-x\theta|^\gamma-1+\gamma x\theta]\beta(\theta)d\theta$$
	and   
	$$J_{\gamma,\epsilon}^2(x)=\int_{|\theta|>\epsilon,|x\theta|>1/2}[|1-x\theta|^\gamma-1]\beta(\theta)d\theta.$$ 
	By Taylor's formula, we get, observing that $|x^2\theta^2|\leq |x|^\gamma|\theta|^\gamma$ if $|x\theta|<1/2$,
	\begin{align*}
	J_{\gamma,\epsilon}^1(x)\leq C\int_{|\theta|>\epsilon,|x\theta|<1/2}x^2\theta^2\beta(\theta)d\theta\leq C|x|^\gamma\int_{-\pi}^{\pi}|\theta|^\gamma\beta(\theta)d\theta.
	\end{align*}
Next, since $|x\theta|>1/2$ implies $1+|x\theta|^\gamma\leq (1+2^\gamma)|x\theta|^\gamma$,
	\begin{align*}
	J_{\gamma,\epsilon}^2(x)\leq C\int_{|\theta|>\epsilon,|x\theta|>1/2}[1+|x\theta|^\gamma]\beta(\theta)d\theta\leq C|x|^\gamma\int_{-\pi}^{\pi}|\theta|^\gamma\beta(\theta)d\theta.
	\end{align*}
We thus have $J_{\gamma,\epsilon}(x)\leq C|x|^\gamma$ and hence $I_{\gamma,\epsilon}(x)\leq C(1+|x|^\gamma$).
\vskip0.5cm

\underline{Step 2}: using Step 1, we now prove
\begin{align*}
\mathbb E\Big[W_\gamma^\gamma(\bar\mu_t^{n,\epsilon},\mu_t^{n,\epsilon})\Big]\leq Ce^{Ct}\int_0^t\mathbb E \big[W_\gamma^\gamma(f_s^\epsilon,\mu_s^{n,\epsilon})\big]ds, 
\end{align*}
for all $t\geq0$. We have $\mathbb E \Big[W_\gamma^\gamma(\bar\mu_t^{n,\epsilon},\mu_t^{n,\epsilon})\Big]\leq \mathbb E(|\Delta_t|^\gamma),$ where $\Delta_t=\bar V_t^1-V_t^1$. It holds
\begin{align*}
\Delta_t=\int_0^t\int_0^1\int_{|\theta|>\epsilon} \Big[(\cos\theta-1)\Delta_{s-}-\sin\theta\delta_{s-}(\alpha)\Big]N^{1}(dsd\alpha d\theta)-b_\epsilon \int_0^t \Delta_sds,
\end{align*}
where $\delta_t(\alpha)=\big( F_t^{n,\epsilon}\big)^{-1}(\alpha)-\big(F_t^\epsilon\big)^{-1}(\alpha)$. By It\^o's formula,
\begin{align*}
\mathbb E\Big[|\Delta_t|^\gamma\Big]&=\mathbb E\Big[\int_0^t\int_0^1\int_{|\theta|>\epsilon} \big[|\cos\theta\Delta_{s}-\sin\theta\delta_{s}(\alpha)|^\gamma-|\Delta_{s}|^\gamma\big]\beta(\theta)dsd\alpha d\theta\Big]\\
&\ \ \ \ -\gamma b_\epsilon\int_0^t\mathbb E(|\Delta_s|^\gamma)ds\\
&\leq\mathbb E\Big[\int_0^t\int_0^1|\Delta_s|^\gamma\int_{|\theta|>\epsilon} \big[|\cos\theta-\sin\theta\frac{\delta_{s}(\alpha)}{\Delta_s}|^\gamma-1\big]\beta(\theta)dsd\alpha d\theta\Big].
\end{align*}
Using Step 1 and then Proposition \ref{Villani}, we get
\begin{align*}
\mathbb E\Big[|\Delta_t|^\gamma\Big]&\leq C\mathbb E\Big[\int_0^t\int_0^1|\Delta_s|^\gamma\Big(1+\frac{|\delta_s(\alpha)|^\gamma}{|\Delta_s|^\gamma}\Big) d\alpha ds\Big]\\
&\leq C\int_0^t\mathbb E[|\Delta_s|^\gamma]ds+C\int_0^t \mathbb E\Big[W_\gamma^\gamma(f_s^\epsilon,\mu_s^{n,\epsilon})\Big]ds.
\end{align*}
We conclude by Gr\"onwall's lemma.
\vskip0.5cm

\underline{Step 3}: using very similar arguments as in the proof of Theorem \ref{systeme} and observing that $W_\gamma(f_t,f_t^\epsilon)\leq W_2(f_t,f_t^\epsilon)$, we get
\begin{align*}
\mathbb E\Big[W_\gamma^\gamma(f_t^\epsilon,\bar \mu_t^{n,\epsilon})\Big]\leq C(1+t)^{\gamma/2}\Big[\epsilon^\gamma+\mathbb E\Big[W_\gamma^\gamma(f_t,\mu_t^n)\Big]\Big],
\end{align*} 
for all $t\geq0$, where $\mu_t^n$ is the empirical measure of a sequence of i.i.d. random variables with law $f_t$.
\vskip0.5cm

\underline{Step 4}: the aim of this step is to prove that
\begin{align} \label{Wgamma}
\sup_{[0,T]}\mathbb E\Big[W_\gamma^\gamma(f_t,\mu_t^{n,\epsilon})\Big]\leq C_T\Big(\epsilon^\gamma+\sup_{[0,T]}\mathbb E\Big[W_\gamma^\gamma(f_t,\mu_t^n)\Big]\Big).
\end{align}
Using the triangular inequality, Theorem \ref{petits sauts}, Step 2 and Step 3, we have
\begin{align*}
\mathbb E\Big[W_\gamma^\gamma(f_t,\mu_t^{n,\epsilon})\Big]&\leq C\Big(W_\gamma^\gamma(f_t,f_t^\epsilon)+\mathbb E\Big[W_\gamma^\gamma(f_t^\epsilon,\bar \mu_t^{n,\epsilon})\Big]+\mathbb E\Big[W_\gamma^\gamma(\bar \mu_t^{n,\epsilon},\mu_t^{n,\epsilon})\Big]\Big)\\
&\leq C_T\Big(\epsilon^\gamma+\mathbb E\Big[W_\gamma^\gamma(f_t,\mu_t^n)\Big]+C_T\int_0^t\mathbb E \Big[W_\gamma^\gamma(f_s^\epsilon,\mu_s^{n,\epsilon})\Big]ds\Big).
\end{align*}
Using again the triangular inequality and Theorem \ref{petits sauts}, we get 
\begin{align*}
\mathbb E \Big[W_\gamma^\gamma(f_s^\epsilon,\mu_s^{n,\epsilon})\Big]&\leq C\Big(\mathbb E \Big[W_\gamma^\gamma(f_s^\epsilon,f_s)\Big]+\mathbb E \Big[W_\gamma^\gamma(f_s,\mu_s^{n,\epsilon})\Big]\Big)\\
&\leq C(1+T)^{\gamma/2}\epsilon^\gamma+C\mathbb E \Big[W_\gamma^\gamma(f_s,\mu_s^{n,\epsilon})\Big].
\end{align*}
We thus have
\begin{align*}
\mathbb E\Big[W_\gamma^\gamma(f_t,\mu_t^{n,\epsilon})\Big]&\leq C_T\Big(\epsilon^\gamma+\mathbb E\Big[W_\gamma^\gamma(f_t,\mu_t^n)\Big]+\int_0^t\mathbb E \Big[W_\gamma^\gamma(f_s,\mu_s^{n,\epsilon})\Big]ds\Big),
\end{align*}
and we conclude with the help of Gr\"onwall's lemma.
\vskip0.5cm

\underline{Step 5}: we can now prove (i). Since $f_0\in\mathcal P_4(\mathbb R)$, Lemma \ref{moment4} implies that $\sup_{t\geq0}\int_{\mathbb R} v^4f_t(dv)<\infty$, whence $\sup_{t\geq0}\mathbb E\Big[W_1(f_t,\mu_t^n)\Big]\leq\frac{C}{\sqrt n}$ by Lemma \ref{emp1}. Inserting this in (\ref{Wgamma}), we easily conclude.
\vskip0.5cm

\underline{Step 6}: we finally prove (ii). Since $f_0\in\mathcal P_p(\mathbb R)$, Lemma \ref{moment4} implies that $\sup_{t\geq0}\int_{\mathbb R} v^pf_t(dv)<\infty$, whence $\sup_{t\geq0}\mathbb E\Big[W_\gamma^\gamma(f_t,\mu_t^n)\Big]\leq\frac{C}{n^{\frac{p-\gamma}{2p-2}}}$ by Lemma \ref{empirique}. Inserting this in (\ref{Wgamma}), we easily conclude. \hfill$\square$

\section{Numerical results}

We consider here the cross section $\beta(\theta)=|\theta|^{-1-\nu}$, with $0<\nu<2$. Let $f_0$  be a probability measure admitting a moment of order 4. We fix an integer $n$, a small parameter $\epsilon>0$, and we take the same notation as in Section \ref{psystem}.

We simulate two systems of particles : the system $(V_t^{i,n,\epsilon})_{t\geq0,i\in\{1,...,n\}}$ described in Section \ref{psystem} (system with diffusion) and the following system without diffusion: for $i\in\{1,...,n\}$ and $t\geq0$, 
\begin{align*}
\tilde V_t^{i,n,\epsilon}=&V_0^{i}+\int_0^t\int_0^1\int_{|\theta|>\epsilon} \Big[(\cos\theta-1)\tilde V_{s-}^{i,n,\epsilon}-\sin\theta \tilde V_{s-}^{j,n,\epsilon}\Big]N^{i,n}(dsd\theta dj).
\end{align*}

The algorithm is the following (we write in italic the parts which only concern the system with diffusion).
\begin{itemize}
	\item We set $t=0$, and for $i=1,...,n$, we simulate $V(i)\sim f_0$ and set $T^{up}(i)=0$.
	\item While $t<T^{final}$ (where $T^{final}$ is the time that we want to reach), we simulate an exponential random variable $T$ with parameter $n\int_{|\theta|>\epsilon}\beta(\theta)d\theta$ and we put $t=t+T$. We choose randomly two integers $i$ and $j$ in $\{1,...,n\}$. \textit{For our system with diffusion, we update the particles $i$ and $j$ by setting}
	\begin{align*}
	 V(i)=V(i)\exp(-b_{\epsilon}(t-T^{up}(i)))+G(i),
	\end{align*}
	\textit{and}
	\begin{align*}
	 V(j)=V(j)\exp(-b_{\epsilon}(t-T^{up}(j)))+G(j),
	\end{align*}
	  \textit{where $G(i)$ (resp. $G(j)$) has a centered Gaussian law with variance $1-\exp(-2b_\epsilon(t-T^{up}(i)))$ (resp. $1-\exp(-2b_\epsilon(t-T^{up}(j)))$), where $b_\epsilon$ is defined in (\ref{ceps}), and we set $T^{up}(i)=T^{up}(j)=t$}.
	  
	  Next, we simulate a random variable $\Theta$ with density $\beta_\epsilon/||\beta_\epsilon||_1$, where $\beta_\epsilon(\theta)=\beta(\theta)\one_{|\theta|>\epsilon}$, and then, for the two systems, we put $V(i)=\cos\Theta V(i)-\sin\Theta V(j)$.
	\item \textit{Only for the system with diffusion, we update all particles with a Gaussian term: for} $i\in\{1,...,n\}$,
	\begin{align*}
	 V(i)=V(i)\exp(-b_{\epsilon}(t-T^{up}(i)))+G(i),
	\end{align*}
	\textit{where $G(i)$ has a centered Gaussian law with variance} $1-\exp(-2b_\epsilon(t-T^{up}(i)))$.
\end{itemize}

For our simulation, we take $T^{final}=0.1$. Our initial data is $f_0=(\delta_{-1}+\delta_1)/2$. The goal here is to see what system is more efficient. For this, we need a reference curve. We obtain it by simulating $n=10^7$ particles with $\epsilon=0.03$, and by using a smoothing procedure. 

\begin{figure}[hbtp]
\includegraphics[width=6cm,height=4cm]{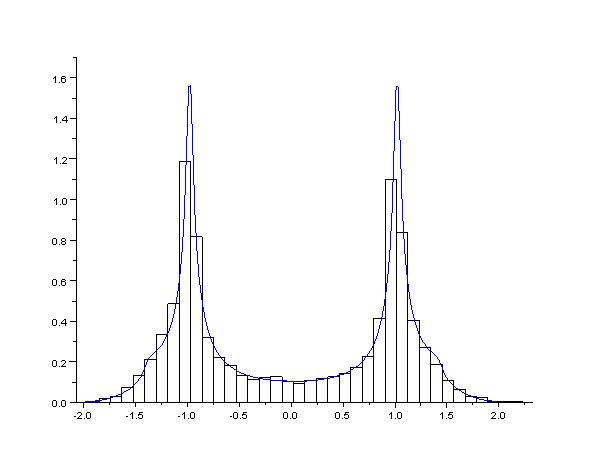}
\includegraphics[width=6cm,height=4cm]{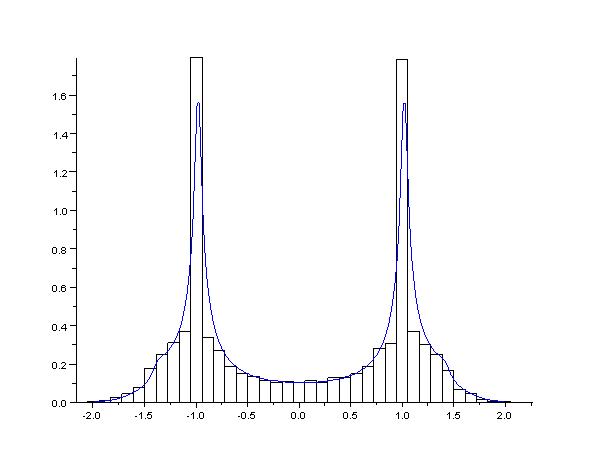}
\caption{$\nu=0.5$. Left graphic: system with diffusion, $n=10^4$, $\epsilon=0.1$. Right graphic: system without diffusion, $n=2.10^4$, $\epsilon=0.1$. Both simulations need approximately 0.05s.}
\end{figure}

\begin{figure}[hbtp]
\includegraphics[width=6cm,height=4cm]{AvecMB0,5_10000_0,1.jpg}
\includegraphics[width=6cm,height=4cm]{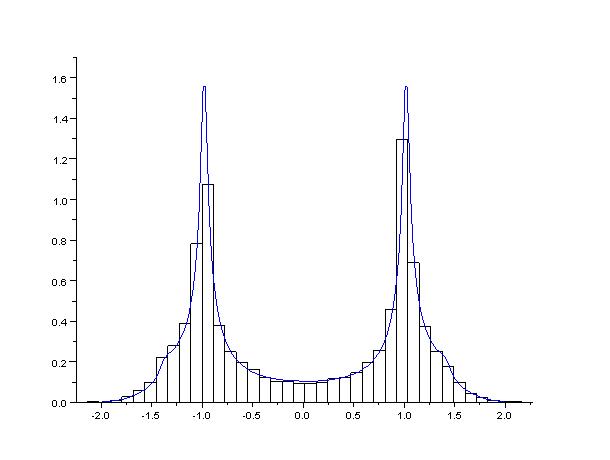}
\caption{$\nu=0.5$. Left graphic: system with diffusion, $n=10^4$, $\epsilon=0.1$. Right graphic: system without diffusion, $n=10^4$, $\epsilon=0.02$. Both simulations need approximately 0.05s.}
\end{figure}

\begin{figure}[hbtp]
\includegraphics[width=6cm,height=4cm]{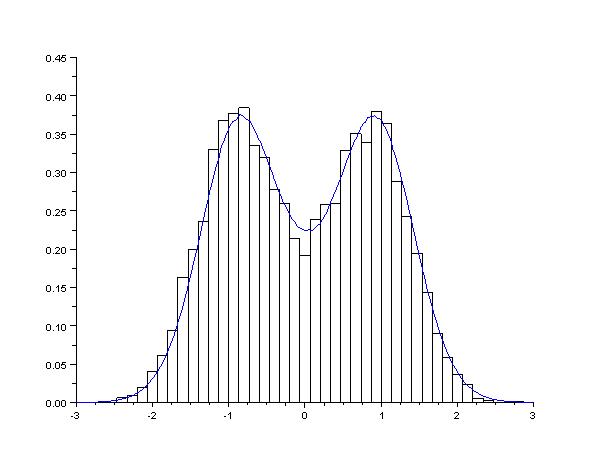}
\includegraphics[width=6cm,height=4cm]{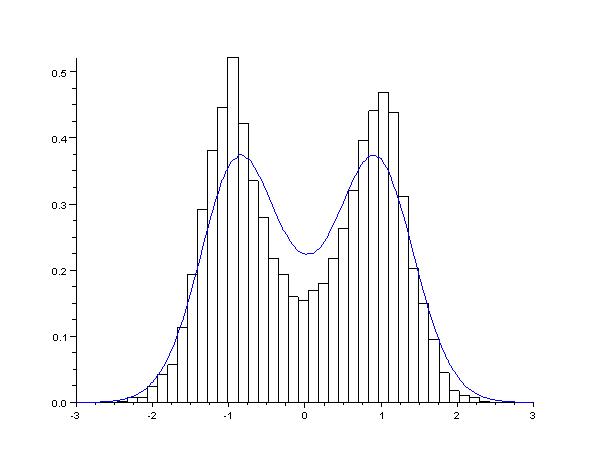}
\caption{$\nu=1.5$. Left graphic: system with diffusion, $n=10^4$, $\epsilon=0.1$. Right graphic: system without diffusion, $n=2.10^4$, $\epsilon=0.1$. Both simulations need approximately 0.14s.}
\end{figure}

\begin{figure}[hbtp]
\includegraphics[width=6cm,height=4cm]{AvecMB1,5_10000_0,1.jpg}
\includegraphics[width=6cm,height=4cm]{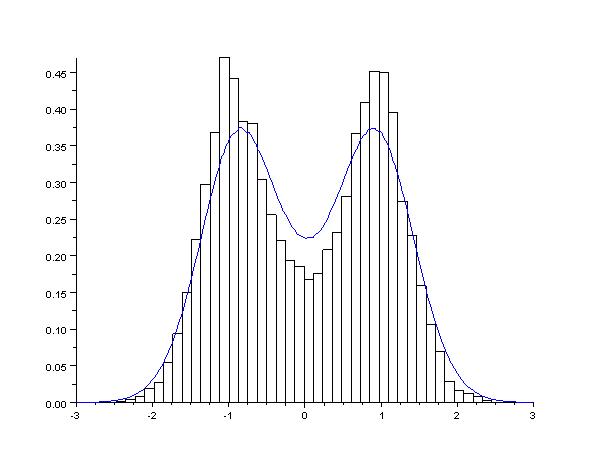}
\caption{$\nu=1.5$. Left graphic: system with diffusion, $n=10^4$, $\epsilon=0.1$. Right graphic: system without diffusion, $n=10^4$, $\epsilon=0.06$. Both simulations need approximately 0.14s.}
\end{figure}

We see that the system with diffusion term is much more efficient when $\nu$ is close to 2. For $\nu$ smaller, the difference not clear.

\appendix

\section{Appendix}

\subsection{Wasserstein distance between a Poisson integral and a Gaussian law}

We start with a result of Rio \cite[Theorem 4.1]{RIO}, which gives some very precise rate of convergence for the standard central limit theorem in Wasserstein distance.
\begin{theo} \label{rio}
There exists a constant $C_0$ such that for any positive integer $n$, for any sequence $(X_i)_{i\geq0}$ of real independant centered random variables in $L^4$,
$$W_2^2(\eta_n,\mathcal{N}(0,1))\leq C_0 v_n^{-2}\sum_{i=1}^n \mathbb E(|X_i|^4),$$
where $\eta_n=\mathcal{L}(v_n^{-1/2}S_n)$, $S_n=\sum_{i=1}^n X_i$, $v_n=Var(S_n)$.
\end{theo}

Using this result, we can estimate the Wasserstein distance between a compensated Poisson integral and a centered Gaussian law with the same variance. The following result is very close to \cite[Corollary 6]{FOU}.   

\begin{coro} \label{coro}
If $E$ is a Polish space endowed with a non-negative $\sigma$-finite measure $\nu$, if N is a Poisson measure on $[0,T]\times E$ with intensity measure $dt\nu(dz)$ and if $H:[0,T]\times E \mapsto \mathbb{R}$ is a deterministic function such that $\int_0^t\int_E (H^2(s,z)+H^4(s,z))\nu(dz)ds<\infty$, then setting 
$$X_t=\int_0^t\int_E H(s,z)\tilde N(ds,dz),\ q_t=\int_0^t\int_E H^2(s,z)\nu(dz)ds,$$
we have
$$W_2^2(\mathcal L(X_t),\mathcal{N}(0,q_t))\leq C_0 \frac{\int_0^t\int_E H^4(s,z)\nu(dz)ds}{q_t},$$
where $C_0$ is a universal constant (the same as in Theorem \ref{rio}).
\end{coro}

\textbf{Proof.}
For $n\geq1$, $i\in\{1,...,n\}$, we set
$$X_i^n=\sqrt{n}\int_{(i-1)t/n}^{it/n}\int_E H(s,z)\tilde N(ds,dz)\quad \text{and}\quad S_n=\sum_{i=1}^n X_i^n.$$
We have $X_t=\frac{S_n}{\sqrt{n}}$. The random variables $X_i^n$ are independent, centered,
$$\mathbb E[(X_i^n)^2]=n\int_{(i-1)t/n}^{it/n}\int_E H^2(s,z)\nu(dz)ds,\quad v_n=Var(S_n)=\sum_{i=1}^n \mathbb E[(X_i^n)^2]=nq_t.$$
It classically holds that
\begin{align*}
\mathbb E[(X_i^n)^4]=n^2\int_{(i-1)t/n}^{it/n}\int_E H^4(s,z)\nu(dz)ds+3n^2\Big(\int_{(i-1)t/n}^{it/n}\int_E H^2(s,z)\nu(dz)ds\Big)^2.
\end{align*}
Hence
\begin{align*}
\sum_{i=1}^n \mathbb E[(X_i^n)^4]=&n^2\int_0^t\int_E H^4(s,z)\nu(dz)ds\\
&+3n^2\sum_{i=1}^n \Big(\int_{(i-1)t/n}^{it/n}\int_E H^2(s,z)\nu(dz)ds\Big)^2.
\end{align*}
By Theorem \ref{rio},
\begin{align*}
W_2^2(\mathcal L(X_t),\mathcal{N}(0,q_t))&=W_2^2\Big(\mathcal L(\frac{1}{\sqrt{n}}S_n),\mathcal{N}(0,q_t)\Big)\\
&=q_tW_2^2\Big(\mathcal L(\frac{1}{\sqrt{v_n}}S_n),\mathcal{N}(0,1)\Big)\\
&\leq C_0\frac{q_t}{v_n^2}	\sum_{i=1}^n \mathbb E[(X_i^n)^4]\\
&\leq C_0\frac{\int_0^t\int_E H^4(s,z)\nu(dz)ds}{q_t}\\
&\quad+3\frac{C_0}{q_t}\sum_{i=1}^n \Big(\int_{(i-1)t/n}^{it/n}\int_E H^2(s,z)\nu(dz)ds\Big)^2.
\end{align*}
Setting $F(t)=\int_0^t|f(s)|ds$ with $f(s)=\int_EH^2(s,z)\nu(dz)$ and observing that $F$ is continuous (and so uniformly continuous on $[0,T]$ for all $T\geq0$), we obtain that $\sum_{i=1}^n \Big(\int_{(i-1)t/n}^{it/n}\int_E H^2(s,z)\nu(dz)ds\Big)^2\rightarrow0$ when $n\rightarrow+\infty$. Since the last formula holds for all $n\geq1$, we easily conclude. \hfill$\square$

\subsection{Rate of convergence of empirical measures}

We first give a classical result about the Wasserstein distance $W_1$.

\begin{lemme}\label{emp1}
Let $\mu$ be a probability measure in $\mathcal P_4(\mathbb R)$. We consider $n$ i.i.d. random variables $(X_i)_{i\in\{1,...,n\}}$ with law $\mu$ and we set $\mu_n=\frac{1}{n}\sum_{i=1}^n \delta_{X_i}$. Then there exists a constant C depending only on $\int_{\mathbb R} x^4\mu(dx)$ such that 
$$\mathbb E\Big[W_1(\mu,\mu_n)\Big]\leq \frac{C}{\sqrt n}.$$
\end{lemme}

\textbf{Proof.}
If we set $F(x)=\mu((-\infty,x])$ and $F_n(x)=\frac{1}{n}\sum_1^n \one_{X_i\leq x}$, we have (see Villani \cite[p 75]{VIL2})
\begin{align*}
\mathbb E\Big(W_1(\mu,\mu_n)\Big)&=\mathbb E\Big(\int_{-\infty}^{\infty}|F(x)-F_n(x)|dx\Big)\\
&=\mathbb E\Big(\int_{-\infty}^{\infty}|F(x)-\frac{1}{n}\sum_{i=1}^n\one_{X_i\leq x}|dx\Big).
\end{align*}
If $Y\sim\mathcal B(n,p)$, $\mathbb E\Big(|\frac{Y}{n}-p|\Big)\leq \sqrt{\mathbb E\Big((\frac{Y}{n}-p)^2\Big)}=\frac{1}{\sqrt n}\sqrt{p(1-p)}$. Hence, since for each $x$, $\sum_{i=1}^n\one_{X_i\leq x}\sim \mathcal B(n,F(x))$,
\begin{align*}
\mathbb E\Big(W_1(\mu,\mu_n)\Big)&\leq\frac{1}{\sqrt n}\int_{-\infty}^{\infty}\sqrt{F(x)(1-F(x))}dx.
\end{align*}
But $A:=\int_{\mathbb R} x^4\mu(dx)<\infty$ implies that for $x\geq1$, $(1-F(x))=\mu([x,+\infty))\leq A/x^4$ and for $x\leq-1$, $F(x)=\mu((-\infty,x])\leq A/x^4$, so that $\int_{-\infty}^{\infty}\sqrt{F(x)(1-F(x))}dx<\infty$.  \hfill$\square$
\vskip0.5cm

We now deduce similar estimates for other Wasserstein distances.

\begin{lemme} \label{empirique}
Let $\mu$ be a probability measure admitting a moment of order $q+\gamma$, with $\gamma>1$ and $q>0$. We consider $n$ i.i.d. random variables $(X_i)_{i\in\{1,...,n\}}$ with law $\mu$ and we set $\mu_n=\frac{1}{n}\sum_{i=1}^n \delta_{X_i}$. There exists a constant C depending on $\gamma$, $q$ and on the moment of $\mu$ of order $q+\gamma$ such that
$$\mathbb E\Big[W_\gamma^\gamma(\mu,\mu_n)\Big]\leq \frac{C}{n^{\frac{q}{2(q+\gamma-1)}}}.$$
\end{lemme}

\textbf{Proof.}
Let us denote by $(\Omega, \mathcal F, \mathbb P)$ the probability space on which $X_1,...,X_n$ are defined. For a fixed $\omega\in\Omega$, we consider two random variables $X$ and $Y^\omega$ defined on the probability space $([0,1],\mathcal B([0,1]),d\alpha)$ with $\mathcal L_\alpha(X)=\mu$ and $\mathcal L_\alpha(Y^\omega)=\mu_n(\omega)$ such that $W_1(\mu,\mu_n(\omega))=\mathbb E_\alpha(|X-Y^\omega|).$ Then we have, for any $A>0$,
$$W_\gamma^\gamma(\mu,\mu_n(\omega))\leq \mathbb E_\alpha(|X-Y^\omega|^\gamma)\leq A^{\gamma-1}W_1(\mu,\mu_n(\omega))+\mathbb E_\alpha(|X-Y^\omega|^\gamma\one_{|X-Y^\omega|>A}).$$
We observe that
\begin{align*}
\mathbb E_\alpha(|X-Y^\omega|^\gamma\one_{|X-Y^\omega|>A})\leq\frac{\mathbb E_\alpha(|X-Y^\omega|^{q+\gamma})}{A^q}.
\end{align*}
But, setting $m_{p}(\mu)=\int_{\mathbb R}|x|^p\mu(dx)$,
\begin{align*}
\mathbb E_\alpha(|X-Y^\omega|^{q+\gamma})\leq C\Big(\mathbb E_\alpha(|X|^{q+\gamma})+\mathbb E_\alpha(|Y^\omega|^{q+\gamma})\Big)=C(m_{q+\gamma}(\mu)+m_{q+\gamma}(\mu_n(\omega)).
\end{align*}
One easily checks that $\mathbb E\Big[m_{q+\gamma}(\mu_n)\Big]=m_{q+\gamma}(\mu)$. Using Lemma \ref{emp1}, we finally get
$$\mathbb E\Big[W_\gamma^\gamma(\mu,\mu_n)\Big]\leq C\Big(\frac{A^{\gamma-1}}{\sqrt n}+\frac{m_{q+\gamma}(\mu)}{A^q}\Big).$$
Choosing $A=n^{\frac{1}{2(q+\gamma-1)}}$ completes the proof. \hfill$\square$

\subsection{Moments of a solution to (\ref{Kac})}

In many places of the proof, we need to upperbound $\int_{\mathbb R} v^4f_t(dv)$ for any $t\geq0$ where $(f_t)_{t\geq0}$ solves (\ref{Kac}). We also need to upperbound higher moments.
\begin{lemme} \label{moment4}
For $f_0\in\mathcal P_4(\mathbb R)$, consider the unique solution $(f_t)_{t\geq0}$ to (\ref{Kac}). For any $t\geq0$, we have 
$$\int_{\mathbb R} v^4f_t(dv)\leq \int_{\mathbb R} v^4f_0(dv)+3\int_{\mathbb R} v^2f_0(dv).$$
If $f_0\in\mathcal P_p(\mathbb R)$ with $p$ even, then there exists a constant $C$ depending on $p$, $\beta$ and on $\int_{\mathbb R}v^p f_0(dv)$ such that for any $t\geq0$,
\begin{align*}
\int_{\mathbb R}v^p f_t(dv)\leq C.
\end{align*}
\end{lemme}

\textbf{Proof.}
We only treat the case $p=4$, see Truesdell \cite{TRU} and Desvillettes \cite{DES3} for the general case.
If we take $\varphi(v)=v^4$, we find, recalling (\ref{Kbeta}) and using that $\beta$ is even,
\begin{align*} 
K_{\beta}^{\varphi}(v,v_{*})= \int_{-\pi}^{\pi}\Big[(\cos^4\theta-1)v^4+\sin^4\theta v_*^4+6\cos^2\theta\sin^2\theta v^2v_*^2\Big]\beta(\theta)d\theta.
\end{align*}
Setting $m_k(\mu)= \int_{\mathbb R}v^k \mu(dv)$ for $\mu$ a probability measure on $\mathbb R$ and $k\in \mathbb N$, we thus get, using (\ref{Kac mesure}),
\begin{align*}
m_4(f_t)=m_4(f_0)+\int_0^t\int_{-\pi}^\pi[&-(1-\cos^4\theta-\sin^4\theta)m_4(f_s)\\
&+6\cos^2\theta\sin^2\theta m_2^2(f_s)]\beta(\theta)d\theta ds.
\end{align*}
Recalling that $m_2(f_s)=m_2(f_0)=:\mathcal E$ for any $s\geq0$, observing that $(\cos^2\theta+\sin^2\theta)^2=1$, which gives $2\cos^2\theta\sin^2\theta=1-\cos^4\theta-\sin^4\theta$ and setting $c=\int_{-\pi}^\pi(1-\cos^4\theta-\sin^4\theta)\beta(\theta)d\theta$, we have
\begin{align*}
m_4(f_t)=m_4(f_0)-c\int_0^tm_4(f_s)ds+3c\mathcal E^2t,
\end{align*}
whence
\begin{align*}
m_4(f_t)=(m_4(f_0)-3\mathcal E^2)\exp(-ct)+3\mathcal E^2\leq m_4(f_0) +3\mathcal E^2,
\end{align*}
as desired. \hfill$\square$

\subsection{Well-posedness for a P.D.E}

To conclude this paper, we state the following result.
\begin{prop} \label{unicite}
For $t\geq0$ and $(v,v_*)\in\mathbb R^2$, we consider two finite non-negative measures $q(t,v,dh)$ and $r(t,v,v_*,dh)$ on $\mathbb R$ such that $\Lambda_q:=\sup_{t,v}q(t,v,\mathbb R)<\infty$, $\Lambda_r:=\sup_{t,v,v_*}r(t,v,v_*,\mathbb R)<\infty$ and for all $T\geq0$, all $(v,v_*)\in\mathbb R^2$,
\begin{align} \label{P2q}
\sup_{[0,T]}\int_{\mathbb R} (h^2+2vh)q(t,v,dh)\leq C_T(1+v^2),
\end{align}
and
\begin{align} \label{P2r}
\sup_{[0,T]}\int_{\mathbb R} (h^2+2vh)r(t,v,v_*,dh)\leq C_T(1+v^2+v_*^2),
\end{align}
Let also $a\geq0$ and $b\in\mathbb R$ be fixed. Then, for any $f_0\in\mathcal P_2(\mathbb R)$, there exists a unique $(f_t)_{t\geq0}\in L_{loc}^\infty([0,\infty),\mathcal P_2(\mathbb R))$ such that for all $\varphi\in C_b^2(\mathbb R)$, all $t\geq0$,
\begin{align} 
\nonumber
\frac{d}{dt}\int_{\mathbb R}\varphi(v) f_t(dv)=&a\int_{\mathbb R}\varphi''(v) f_t(dv)+b\int_{\mathbb R}v\varphi'(v) f_t(dv)\\
\label{PDE}
&+\int_{\mathbb R}\int_{\mathbb R}\Big[\varphi(v+h)-\varphi(v)\Big]q(t,v,dh) f_t(dv)\\
\nonumber
&+\int_{\mathbb R}\int_{\mathbb R}\int_{\mathbb R}\Big[\varphi(v+h)-\varphi(v)\Big]r(t,v,v_*,dh) f_t(dv)f_t(dv_*).
\end{align}
\end{prop}

\textbf{Proof.}
We denote by $\mathcal M(\mathbb R)$ the set of finite signed measures on $\mathbb R$. If $\mu\in\mathcal M(\mathbb R)$, we set $|\mu|_{TV}=\sup_{\varphi\in L^\infty, ||\varphi||_\infty\leq1}\int_{\mathbb R}\varphi(v)\mu(dv)$. Using the Lusin Theorem (see e.g. \cite[Theorem 9.11]{BRI}), we have $|\mu|_{TV}=\sup_{\varphi\in C_b, ||\varphi||_\infty\leq1}\int_{\mathbb R}\varphi(v)\mu(dv)$. We also have $|\mu|_{TV}=\int_{\mathbb R}|\mu|(dv)$ where $|\mu|=\mu_+ +\mu_-$ and if $\mu$ has a density $f$ with respect to the Lebesgue measure, $|\mu|_{TV}=\int_{\mathbb R}|f(v)|dv$. 
\vskip0.5cm

\textit{Preliminaries.}
For $\epsilon>0$, we set $G_\epsilon(v)=\frac{1}{\sqrt{2\pi\epsilon}}e^{\frac{-v^2}{2\epsilon}}$. Let $\mu\in\mathcal M(\mathbb R)$. We claim that $\lim_{\epsilon\rightarrow0}|\mu*G_\epsilon|_{TV}=|\mu|_{TV}$. Observe that this is not obvious, since it does not hold true, generally, that $\lim_{\epsilon\rightarrow0}|\mu*G_\epsilon-\mu|_{TV}=0$ (choose e.g. $\mu=\delta_0$). First, we have
\begin{align*}
|\mu*G_\epsilon|_{TV}=\int_{\mathbb R}|\mu*G_\epsilon(v)|dv&=\int_{\mathbb R}\Big|\int_{\mathbb R}G_\epsilon(v-w)\mu(dw)\Big|dv\\ 
&\leq\int_{\mathbb R}\int_{\mathbb R}G_\epsilon(v-w)dv|\mu(dw)|=|\mu|_{TV}.
\end{align*}
Next, let $\alpha>0$. There exists a function $\varphi\in C_b$ with $||\varphi||_\infty\leq1$ such that $\int_{\mathbb R}\varphi(v)\mu(dv)\geq|\mu|_{TV}-\alpha$. We have, since $\mu*G_\epsilon$ clearly converges weakly (in the sence of measures) to $\mu$,
\begin{align*}
|\mu*G_\epsilon|_{TV}\geq\int_{\mathbb R}\varphi(v)(\mu*G_\epsilon)(v)dv\stackrel{\epsilon\rightarrow0}{\longrightarrow}\int_{\mathbb R}\varphi(v)\mu(dv)\geq|\mu|_{TV}-\alpha.
\end{align*}
Making $\alpha$ tend to zero, we get $\liminf_{\epsilon\rightarrow0}|\mu*G_\epsilon|_{TV}\geq|\mu|_{TV}$. 
\vskip0.5cm

\textit{Uniqueness.}
We consider two solutions $(f_t)_{t\geq0}$ and $(g_t)_{t\geq0}$, with $f_0=g_0$ and for $t\geq0$ we set $\mu_t=f_t-g_t$. For any $\varphi\in C_b^2(\mathbb R)$, any $t\geq0$
\begin{align*}
\frac{d}{dt}\int_{\mathbb R}\varphi(v) \mu_t(dv)=&a\int_{\mathbb R}\varphi''(v) \mu_t(dv)+b\int_{\mathbb R}v\varphi'(v) \mu_t(dv)\\
&+\int_{\mathbb R}\int_{\mathbb R}\Big[\varphi(v+h)-\varphi(v)\Big]q(t,v,dh) \mu_t(dv)\\
&+\int_{\mathbb R}\int_{\mathbb R}\int_{\mathbb R}\Big[\varphi(v+h)-\varphi(v)\Big]r(t,v,v_*,dh)[f_t(dv)f_t(dv_*)\\
&\quad\quad\quad\quad\quad\quad\quad\quad\quad\quad\quad\quad\quad\quad\quad\quad\quad\quad-g_t(dv)g_t(dv_*)].
\end{align*}
We first observe that $f_t(dv)f_t(dv_*)-g_t(dv)g_t(dv_*)=f_t(dv)\mu_t(dv_*)+g_t(dv_*)\mu_t(dv)$. We have
\begin{align*}
\partial_t(\mu_t*G_\epsilon)(v)&=\frac{d}{dt}\int_{\mathbb R}G_\epsilon(v-w)\mu_t(dw)\\
&=a\int_{\mathbb R}G_\epsilon''(v-w)\mu_t(dw)-b\int_{\mathbb R}wG_\epsilon'(v-w)\mu_t(dw)\\
&\quad+\int_{\mathbb R}\int_{\mathbb R}[G_\epsilon(v-w-h)-G_\epsilon(v-w)]\mu_t(dw)q(t,w,dh)\\
&\quad+\int_{\mathbb R}\int_{\mathbb R}\int_{\mathbb R}[G_\epsilon(v-w-h)-G_\epsilon(v-w)][f_t(dw)\mu_t(dw_*)\\
&\quad\quad\quad\quad\quad\quad\quad\quad\quad\quad\quad\quad\quad\quad\quad+g_t(dw_*)\mu_t(dw)]r(t,w,w_*,dh).
\end{align*}
For $\eta>0$, we consider a function $\Gamma_\eta$ of class $C^2$ such that for any $x\in\mathbb R$, $(|x|-\eta)_+\leq\Gamma_\eta(x)\leq|x|$, $\Gamma_\eta''(x)\geq0$ and $||\Gamma_\eta'||_\infty\leq1$. We also assume that $\Gamma_{\eta_1}\geq\Gamma_{\eta_2}$ if $\eta_1\leq\eta_2$. Observing that $\int_{\mathbb R}G_\epsilon''(v-w)\mu_t(dw)=(\mu_t*G_\epsilon)''(v)$, we have
\begin{align*}
\frac{d}{dt}\int_{\mathbb R}\Gamma_\eta((\mu_t*G_\epsilon)(v))dv&=\int_{\mathbb R}\Gamma_\eta'((\mu_t*G_\epsilon)(v))\partial_t(\mu_t*G_\epsilon)(v)dv\\
&=A_t+B_t+C_t+D_t,
\end{align*}
where
\begin{align*}
A_t=a\int_{\mathbb R}\Gamma_\eta'((\mu_t*G_\epsilon)(v))(\mu_t*G_\epsilon)''(v)dv,
\end{align*}
\begin{align*}
B_t=-b\int_{\mathbb R}\int_{\mathbb R}\Gamma_\eta'((\mu_t*G_\epsilon)(v))wG_\epsilon'(v-w)\mu_t(dw)dv,
\end{align*}
\begin{align*}
C_t=\int_{\mathbb R}\int_{\mathbb R}\int_{\mathbb R}\Gamma_\eta'((\mu_t*G_\epsilon)(v))[G_\epsilon(v-w-h)-G_\epsilon(v-w)]\mu_t(dw)q(t,w,dh)dv,
\end{align*}
and
\begin{align*}
D_t=\int_{\mathbb R}\int_{\mathbb R}\int_{\mathbb R}\int_{\mathbb R}\Gamma_\eta'((\mu_t*G_\epsilon)(v))[G_\epsilon &(v-w-h)-G_\epsilon(v-w)][f_t(dw)\mu_t(dw_*)\\
&+g_t(dw_*)\mu_t(dw)]r(t,w,w_*,dh)dv.
\end{align*}
Using an integration by parts and recalling that $\Gamma_\eta''(x)\geq0$ for any $x\in\mathbb R$, we have
\begin{align*}
A_t= -a\int_{\mathbb R}\Gamma_\eta''((\mu_t*G_\epsilon)(v))\big((\mu_t*G_\epsilon)'(v)\big)^2dv\leq0.
\end{align*}
First writing $w=v+w-v$ and then using an integration by parts (observe that $\int_{\mathbb R}G_\epsilon'(v-w)\mu_t(dw)=(\mu_t*G_\epsilon)'(v)$), we have
\begin{align*}
B_t&\leq -b\int_{\mathbb R}\int_{\mathbb R}\Gamma_\eta'((\mu_t*G_\epsilon)(v))vG_\epsilon'(v-w)\mu_t(dw)dv\\
&\quad+||\Gamma_\eta'||_\infty|b|\int_{\mathbb R}\int_{\mathbb R}|w-v||G_\epsilon'(v-w)||\mu_t(dw)|dv\\
&\leq b\int_{\mathbb R}\Gamma_\eta((\mu_t*G_\epsilon)(v))dv+|b|\int_{\mathbb R}|v||G_\epsilon'(v)|dv\int_{\mathbb R}|\mu_t|(dw)\\
&\leq C\Big(\int_{\mathbb R}|(\mu_t*G_\epsilon)(v)|dv+|\mu_t|_{TV}\Big)\leq C|\mu_t|_{TV}.
\end{align*}
We used the preliminaries and the fact that $\int_{\mathbb R}|v||G_\epsilon'(v)|dv\leq C$. Using next that $\int_{\mathbb R}G_\epsilon(v-w-h)dv=\int_{\mathbb R}G_\epsilon(v-w)dv=1$, we have
\begin{align*}
C_t+D_t&\leq 2||\Gamma_\eta'||_\infty\Lambda_q\int_{\mathbb R}|\mu_t|(dw)+2||\Gamma_\eta'||_\infty\Lambda_r\Big(\int_{\mathbb R}|\mu_t|(dw_*)+\int_{\mathbb R}|\mu_t|(dw)\Big)\\
&\leq C|\mu_t|_{TV}.
\end{align*}
We thus get
\begin{align*}
\frac{d}{dt}\int_{\mathbb R}\Gamma_\eta(\mu_t*G_\epsilon(v))dv\leq C|\mu_t|_{TV}.
\end{align*}
Using the monotone convergence Theorem (recall that $\Gamma_\eta(x)$ increases to $|x|$ as $\eta$ decreases to 0) and recalling that $\mu_0=0$, we have
\begin{align*}
|\mu_t*G_\epsilon|_{TV}&=\lim_{\eta\rightarrow0}\int_{\mathbb R}\Gamma_\eta((\mu_t*G_\epsilon)(v))dv\\
&\leq\lim_{\eta\rightarrow0}\int_{\mathbb R}\Gamma_\eta((\mu_0*G_\epsilon)(v))dv+C\int_0^t|\mu_s|_{TV}ds\\
&\leq C\int_0^t|\mu_s|_{TV}ds.
\end{align*}
Making $\epsilon$ tend to 0 and using the preliminaries, we get,
\begin{align*}
|\mu_t|_{TV}&\leq C\int_0^t|\mu_s|_{TV}ds,
\end{align*}
and we deduce that $|\mu_t|_{TV}=0$ by Gr\"onwall's lemma.
\vskip0.5cm

\textit{Existence.}
For $(Q_t)_{t\geq0}\in L_{loc}^\infty([0,\infty),\mathcal P_2(\mathbb R))$ given, we consider the following linear P.D.E. with unknown $(g_t^Q)_{t\geq0}$: for all $\varphi\in C_b^2(\mathbb R)$, all $t\geq0$,
\begin{align}
\nonumber
\frac{d}{dt}\int_{\mathbb R}\varphi(v) g_t^Q(dv)=&a\int_{\mathbb R}\varphi''(v) g_t^Q(dv)+b\int_{\mathbb R}v\varphi'(v) g_t^Q(dv)\\
\label{PDElinear}
&+\int_{\mathbb R}\int_{\mathbb R}\Big[\varphi(v+h)-\varphi(v)\Big]q(t,v,dh) g_t^Q(dv)\\
\nonumber
&+\int_{\mathbb R}\int_{\mathbb R}\int_{\mathbb R}\Big[\varphi(v+h)-\varphi(v)\Big]r(t,v,v_*,dh) g_t^Q(dv)Q_t(dv_*).
\end{align}
For $t\geq0$ and $(v,v_*)\in\mathbb R^2$, we consider the following probability measures
\begin{align*}
\eta_{t,v}^q(dh):=\frac{q(t,v,dh)}{\Lambda_q}+\Big(1-\frac{q(t,v,\mathbb R)}{\Lambda_q}\Big)\delta_0(dh)
\end{align*}
and 
\begin{align*}
\eta_{t,v,v_*}^r(dh):=\frac{r(t,v,v_*,dh)}{\Lambda_r}+\Big(1-\frac{r(t,v,v_*,\mathbb R)}{\Lambda_r}\Big)\delta_0(dh),
\end{align*}
and we set $F_{t,v}^q(x):=\eta_{t,v}^q((-\infty,x])$ and $F_{t,v,v_*}^r(x):=\eta_{t,v,v_*}^r((-\infty,x])$. We also set $H^q(t,v,u):=(F_{t,v}^q)^{-1}(u)$, $H^r(t,v,v_*u):=(F_{t,v,v_*}^r)^{-1}(u)$ and we consider the following S.D.E.
\begin{align}
\nonumber
V_t=&V_0+\int_0^t\int_0^1 H^q(s,V_{s-},u) N^q(ds du)+\int_0^t\int_{\mathbb R}\int_0^1H^r(s,V_{s-},v_*,u)N^r(dsdv_*du)\\
\label{annex}
&+b\int_0^t V_s ds + \sqrt {2a} B_t, 
\end{align}
where $N^q$ is a Poisson measure with intensity measure $\Lambda_q dsdu$, $N^r$ is a Poisson measure with intensity measure $\Lambda_r dsduQ_s(dv_*)$ and $B$ is a Brownian motion. There is existence and uniqueness for this S.D.E. because the Poisson measures $N^q$ and $N^r$ are finite, and because the drift and diffusion coefficients are Lipshitz-continuous (see Ikeda-Watanabe \cite{IKE}). Using It\^o's formula and taking expectations, we get, for any $\varphi\in C_b^2(\mathbb R)$,
\begin{align*}
\mathbb E[\varphi(V_t)]&=\mathbb E[\varphi(V_0)]+ \int_0^t\mathbb E\Big[\int_0^1[\varphi(V_s+H^q(s,V_{s},u))-\varphi(V_s)]\Lambda_q du\Big]ds\\
&\quad +\int_0^t\mathbb E\Big[\int_{\mathbb R}\int_0^1[\varphi(V_s+H^r(s,V_{s},v_*,u))-\varphi(V_s)]\Lambda_r Q_s(dv_*)du\Big]ds\\
&\quad +b\int_0^t\mathbb E[\varphi'(V_s)V_s]ds+a\int_0^t\mathbb E[\varphi''(V_s)]ds\\
&=\mathbb E[\varphi(V_0)]+ \int_0^t\mathbb E\Big[\int_{\mathbb R}[\varphi(V_s+h)-\varphi(V_s)]q(s,V_s,dh)\Big] ds \\
&\quad +\int_0^t\mathbb E\Big[\int_{\mathbb R}\int_{\mathbb R}[\varphi(V_s+h)-\varphi(V_s)]r(s,V_s,v_*,dh)Q_s(dv_*)\Big]ds\\
&\quad +b\int_0^t\mathbb E[\varphi'(V_s)V_s]ds+a\int_0^t\mathbb E[\varphi''(V_s)]ds.
\end{align*}
Setting $g_t^Q=\mathcal L(V_t)$, we thus realize that $(g_t^Q)_{t\geq0}$ solves (\ref{PDElinear}). 

If $(Q_t)_{t\geq0}$ and $(R_t)_{t\geq0}$ are in $L_{loc}^\infty([0,\infty),\mathcal P_2(\mathbb R))$, then by the same kind of arguments as in the uniqueness proof, we have for any $t\in[0,T]$, denoting $\mu_t=g_t^Q-g_t^R$,
\begin{align*}
|\mu_t|_{TV}\leq C\int_0^t(|\mu_s|_{TV}+|Q_s-R_s|_{TV})ds,
\end{align*}
whence by Gr\"onwall's Lemma,
\begin{align*}
\sup_{[0,T]}|\mu_t|_{TV}\leq C_T\int_0^T|Q_s-R_s|_{TV}ds.
\end{align*}
We consider $f_0\in\mathcal P_2(\mathbb R)$. For $t\geq0$, we set $f_t^0=f_0$ and $f_t^{k+1}=f_t^{f^k}$. Then we have
\begin{align*}
\sup_{[0,T]}|f_t^{k+1}-f_t^k|_{TV}&\leq C_T\int_0^T|f_s^k-f_s^{k-1}|_{TV}ds.
\end{align*}
We classically conclude that $(f_t^k)_{t\geq0}$ converges as $k$ tends to infinity to some $(f_t)_{t\geq0}$ solving (\ref{PDE}). Using (\ref{P2q}), (\ref{P2r}) and (\ref{PDE}) with $\varphi(v)=v^2$, we see that $(f_t)_{t\geq0}\in L_{loc}^\infty([0,\infty),\mathcal P_2(\mathbb R))$.

\end{document}